\documentclass[12pt]{article}
\usepackage{amssymb}
\usepackage{amsmath}
\usepackage{graphicx,color}
\usepackage{epstopdf}

\newcommand{\bea}{\begin{eqnarray}}
\newcommand{\eea}{\end{eqnarray}}

\numberwithin{equation}{section}

\begin{document}
\begin{titlepage}
%
%
\vspace*{10mm}
\begin{center}
\baselineskip 25pt 
{\Large\bf
Electroweak vacuum stability in classically conformal $B-L$ extension of the Standard Model
}
\end{center}
\vspace{5mm}
\begin{center}
{\large
Arindam Das\footnote{adas8@ua.edu},
Nobuchika Okada\footnote{okadan@ua.edu}
and
Nathan Papapietro\footnote{npapapietro@ua.edu}
}
\end{center}
\vspace{2mm}

\begin{center}
{\it
Department of Physics and Astronomy, University of Alabama, \\
Tuscaloosa, Alabama 35487, USA
}
\end{center}
\vspace{0.5cm}
\begin{abstract}

We consider the minimal U(1)$_{B-L}$ extension of the Standard Model (SM) with the classically conformal invariance,   
   where an anomaly free U(1)$_{B-L}$ gauge symmetry is introduced 
   along with three generations of right-handed neutrinos and a U(1)$_{B-L}$ Higgs field. 
Because of the classically conformal symmetry, all dimensional parameters are forbidden. 
   The $B-L$ gauge symmetry is radiatively broken through the Coleman-Weinberg mechanism, 
   generating the mass for the $U(1)_{B-L}$ gauge boson ($Z^\prime$ boson) and 
   the right-handed neutrinos. 
Through a small negative coupling between the SM Higgs doublet and the $B-L$ Higgs field, 
   the negative mass term for the SM Higgs doublet is generated and the electroweak symmetry is broken.  
In this model context, we investigate the electroweak vacuum instability problem in the SM. 
It is known that in the classically conformal U(1)$_{B-L}$ extension of the SM, 
  the electroweak vacuum remains unstable in the renormalization group analysis at the one-loop level.   
In this paper, we extend the analysis to the two-loop level, and perform parameter scans. 
We identify a parameter region which not only solve the vacuum instability problem, 
   but also satisfy the recent ATLAS and CMS bounds from search for $Z^\prime$ boson resonance 
   at the LHC Run-2. 
Considering self-energy corrections to the SM Higgs doublet through the right-handed neutrinos 
  and the $Z^\prime$ boson, we derive the naturalness bound on the model parameters 
  to realize the electroweak scale without fine-tunings. 
  
\end{abstract}
\end{titlepage}

\section{Introduction}
The stability of the electroweak scale is one of the biggest mysteries in the Standard Model (SM), 
   since the self-energy of the SM Higgs doublet field receives quantum corrections 
   which are quadratically sensitive to the ultraviolet cutoff of the SM. 
A fine-tuning of the Higgs mass parameter is required to reproduce the correct electroweak scale 
   if the ultraviolet cutoff scale is far above the electroweak scale (the gauge hierarchy problem).  
This problem can be solved if new physics beyond the SM makes the self-energy   
   of the SM Higgs doublet insensitive (or logarithmically sensitive) to the ultraviolet cutoff. 
It is well-known that supersymmetric extension of the SM  can achieve this insensitivity. 
Despite lots of efforts of searching for supersymmetry at the Large Hadron Collider (LHC) experiments, 
   current LHC data include less indications for productions of supersymmetric particles. 
Hence we may seek other possibilities to solve the gauge hierarchy problem without supersymmetry.

According to the argument by Bardeen \cite{Bardeen}
   once the classical conformal invariance and its minimal violation by quantum anomalies are imposed 
   on the SM (or the general Higgs model), the model can be logarithmically sensitive to the ultraviolet cutoff. 
If this is the case, we introduce the classically conformal symmetry to the SM  
    to make the model free from the quadratic corrections.\footnote{ 
In terms of the ultraviolet completion, one may consider a conformal model into which the SM is embedded. 
Based on a toy model, it has been shown in \cite{Tavares:2013dga} that the SM Higgs mass 
  is sensitive to a scale at which the SM merges into a conformal field theory.   
Since conformal field theories in 4-dimensions have not yet been completely understood,  
  it is highly non-trivial to verify if this sensitivity is inevitable. 
Hence, we leave this issue in this paper and assume that the SM Higgs does not receive 
  quadratic corrections to its self energy. }     
In this system, there is no mass parameter in the original Lagrangian, 
   and the mass scale must be generated by quantum corrections. 
The massless U(1) Higgs model discussed by Coleman and Weinberg \cite{CW} nicely fits this picture, 
   where the model is defined as a massless, conformal invariant theory,  
   and the U(1) gauge symmetry is radiatively broken by the Coleman-Weinberg (CW) mechanism, 
   generating a mass scale through the dimensional transmutation.

Recently, the extension of the SM with the classically conformal invariance has received a fair amount of attention,  
   and many models in this direction have been proposed \cite{models}-\cite{OOT}. 
Among them, the classically conformal U(1)$_{B-L}$ extension of the SM \cite{IOO1, IOO2} 
   is a very simple and well-motivated model, 
   since the $B-L$ (baryon number minus lepton number) is a unique anomaly-free  global symmetry 
   and it can be easily gauged. 
Once the U(1)$_{B-L}$ is gauged, we need new chiral fermions to cancel the U(1)$_{B-L}$ gauge 
   and the mixed gravitational anomalies. 
The simplest possibility is to introduce three right-handed neutrinos, which are nothing but the particles 
   that we need to incorporate the neutrino mass in the SM.  
In this conformal symmetric model, the $B-L$ gauge symmetry is broken by the vacuum expectation value (VEV) 
   of the $B-L$ Higgs field developed by the CW mechanism, 
   and the masses for $Z^\prime$ boson and three right-handed neutrinos are generated. 
This radiative $B-L$ gauge symmetry breaking is the sole origin of mass scale in this model, 
  and the negative mass squared for the SM Higgs doublet is generated by this symmetry breaking \cite{IOO1}.

The SM Higgs boson is finally discovered at the LHC, and the experimental confirmations of  
   the Higgs properties in the SM has just begun. 
According to the SM, we can read off the value of the quartic Higgs coupling at the electroweak scale 
  from the measured Higgs boson mass, and we can investigate the behavior of the Higgs potential 
  toward high energies by extrapolating the quartic coupling through its renormalization group evolution. 
It turns out that the running quartic coupling becomes negative around $10^{10}$ GeV \cite{RGEs}, and this fact means 
  that the electroweak vacuum is not stable.   
Practically, this instability may not be a problem, since the lifetime of our electroweak vacuum 
  is estimated to be much longer than the age of the universe \cite{Metastable}. 
However, in our context of the classically conformal extension of the SM, this electroweak vacuum instability 
  seems to cause a theoretical inconsistency. 
The instability indicates that the electroweak symmetry is radiatively broken at a very high energy,  
  which in turn generates a large mass term for the $B-L$ Higgs field. 
Therefore, with such a large mass, the $B-L$ symmetry breaking is no longer trigged by the CW mechanism.

In this paper, we investigate the electroweak vacuum stability in the context of the classically conformal 
   U(1)$_{B-L}$  extension of the SM.  
It is known that the electroweak vacuum is still unstable in this context in the renormalization group 
   analysis at the one-loop level~\cite{Khoze, OOT}. 
We extend the analysis to the two-loop level and find that there exist parameter regions 
   which can keep the electroweak vacuum stable.    
In our analysis, we use the result from the combined analysis by the ATLAS and the CMS experiments 
   for the Higgs boson mass measurement in the range of 
    $m_{h}=$125.09$\pm$ 0.21 (stat.)$\pm$ 0.11 (syst.) GeV~\cite{Higgs1} and 
   the recent  result of top quark mass measurement $m_{t}=$ 172.38$\pm$ 0.10$\pm$ 0.65~\cite{top1} by the CMS experiments. 
We also consider the current collider bounds, namely, 
   a lower bound on the $B-L$ gauge symmetry breaking scale from the LEP electroweak precision measurements,  
   and a lower bound on the $Z^\prime$ boson mass from the recent ATLAS~\cite{ATLAS13} and CMS~\cite{CMS13} results at the LHC Run-2.    
In addition, we evaluate self-energy corrections to the SM Higgs doublet from the heavy states, 
  the $Z^\prime$ boson and the right-handed neutrinos associated with the $B-L$ symmetry breaking, 
  and find naturalness bounds to reproduce the electroweak scale without any fine-tunings of model parameters.

This paper is organized as follows. 
Our model is defined in the next section. 
In Sec.~3, we discuss the radiative $B-L$ symmetry breaking through the CW mechanism 
  and the electroweak symmetry breaking triggered by it.   
In Sec.~4, we analyze the renormalization group evolutions of the couplings at the two-loop level, 
  and find a parameter regions which can keep the quartic SM Higgs coupling to be positive 
  anywhere between the electroweak scale and the Planck scale. 
We also consider the current collider bounds of the model parameters, 
  in particular, the recent ATLAS and CMS results of search for $Z^\prime$ boson resonance 
  at the LHC Run-2 are interpreted to our $B-L$ model. 
In Sec.~5, we evaluate self-energy corrections to the SM Higgs doublet, and derive 
  the naturalness bounds to reproduce the electroweak scale without fine-tunings for the model parameters. 
We summarize our results in Sec.~6. 
Formulas we used in our analysis are listed in Appendices.

\section{Classically conformal $U(1)_{B-L}$ extended SM }
{\renewcommand{\arraystretch}{1.5} 
\begin{table}[ht]
\begin{center}
\begin{tabular}{c| c c|c c}
      &  SU(3)$_C$  & SU(2)$_L$ & U(1)$_Y$ & U(1)$_{B-L}$ \\ 
\hline
$q^{i}_{L}$ & {\bf 3 }    &  {\bf 2}         & $\frac{1}{6}$       & $\frac{1}{3}$  \\
$u^{i}_{R}$ & {\bf 3 }    &  {\bf 1}         & $\frac{2}{3}$       & $\frac{1}{3}$  \\
$d^{i}_{R}$ & {\bf 3 }    &  {\bf 2}         & $-\frac{1}{3}$       & $\frac{1}{3}$  \\
\hline
$\ell^{i}_{L}$ & {\bf 1 }    &  {\bf 2}         & $-\frac{1}{2}$       & $-1$  \\
$e^{i}_{R}$    & {\bf 1 }    &  {\bf 1}         & $-1$                   & $-1$  \\
\hline
$H$            & {\bf 1 }    &  {\bf 2}         & $-\frac{1}{2}$       & $0$  \\
\hline
$\Phi$         & {\bf 1 }    &  {\bf 1}         &$ 0$                    & $+2$  \\
$N^{j}_{R}$    & {\bf 1 }    &  {\bf 1}         &$0$                    & $-1$  \\
\hline
\end{tabular}
\end{center}
\caption{
The particle contents of the $U(1)_{B-L}$ extended SM. 
In addition to the SM particle contents, the right-handed neutrino 
   $N_R^i$ ($i=1,2,3$ denotes the generation index) and a complex scalar $\Phi$ are introduced. 
}
\label{table1}
\end{table}

We investigate the minimal U(1)$_{B-L}$ extension of the SM with the classically conformal invariance, 
  where the model is based on the gauge group 
  SU(3)$_C \times$SU(2)$_L \times$U(1)$_Y \times$U(1)$_{B-L}$. 
The particle contents of the model are listed in Table~\ref{table1}. 
In addition to the SM particle contents, we introduce the $B-L$ Higgs field with the $B-L$ charge $2$ ($\Phi$) 
   and three right-handed neutrinos ($N_R^i$) for cancelation of all the gauge and gravitational anomalies.  
The covariant derivative relevant to U(1)$_Y \times$ U(1)$_{B-L}$ is given by 
\begin{eqnarray}
  D_\mu = \partial_\mu - i (Q_Y~~Q_{BL}) 
	              \left( \begin{array}{cc}
				g_{1} & g_{YB} \\	
				g_{BY} & g_{BL} 
		     \end{array} \right)
	\left( \begin{array}{c} 
	   B_\mu \\ 
	    Z^\prime_\mu
     \end{array} \right), 
\label{Dmu}
\end{eqnarray}
where $Q_Y$ and $Q_{BL}$ are U(1)$_Y$ and U(1)$_{B-L}$ charges of a particle, respectively,  
   and $g$s are the gauge couplings.  
Because of  the kinetic mixing between the two U(1) gauge bosons, the off-diagonal elements 
  ($g_{YB}$ and $g_{BY}$) are introduced. 
In the following analysis, we take the boundary condition, $g_{YB}=g_{BY}=0$,  
   at the $B-L$ symmetry breaking scale, where the two U(1) gauge bosons are 
   diagonal with each other, for simplicity.

The Yukawa sector of the SM is extended to have 
\bea
\mathcal{L}_{Yukawa} \supset  -Y^{ij}_{D} \overline{\ell^i_{L}} H N_R^j -\frac{1}{2} Y^k_{N} \Phi \overline{N_R^{k C}} N_R^k 
  + {\rm h.c.} ,
\label{Lag1} 
\eea
where the first term is the neutrino Dirac Yukawa coupling, while the second term is the Majorana Yukawa coupling. 
Without loss of generality, we have already diagonalized the Majorana Yukawa coupling.   
The $B-L$ gauge symmetry breaking generates the Majorana neutrino mass term in the second term. 
The seesaw mechanism~\cite{seesaw} is automatically implemented in the model after the electroweak symmetry breaking.

We apply the classically conformal invariance to the model, and the scalar potential is given by
\bea
V = \lambda (H^{\dagger}H)^{2} + \lambda_{2} (\Phi^{\dagger}\Phi)^{2} + \lambda_{3} (H^{\dagger}H) (\Phi^{\dagger}\Phi). 
\label{Pot1}
\eea
Note that the mass terms are all forbidden by the conformal invariance. 
If  $\lambda_3$ is negligibly small, we can analyze the Higgs potential separately for $\Phi$ and $H$. 
This will be justified in the next section. 
When the Majorana Yukawa coupling $Y_N^i$ is negligible compared to the U(1)$_{B-L}$ gauge coupling, 
 the $\Phi$ sector is identical with the original Coleman-Weinberg model \cite{CW}, 
   so that the U(1)$_{B-L}$ gauge symmetry is radiatively broken. 
The mass term for the SM Higgs doublet is generated through $\lambda_3$ with the non-zero VEV of $\Phi$,  
   and the electroweak symmetry is broken when we choose $\lambda_3 <0$~\cite{IOO1}. 
Therefore,  the electroweak symmetry breaking is driven by the radiative $B-L$ symmetry breaking.

\section{Radiative gauge symmetry breakings }
Assuming a negligibly small $\lambda_3$, we first analyze the U(1)$_{B-L}$ Higgs sector. 
Without mass terms, the CW potential \cite{CW} at the one-loop level (in the Landau gauge) 
   is found to be
\begin{eqnarray}
  V(\phi) =  \frac{\lambda_2}{4} \phi^4 
 + \frac{\beta_\Phi}{8} \phi^4 \left(  \ln \left[ \frac{\phi^2}{M^2} \right] - \frac{25}{6} \right), 
\label{Eq:CW_potential} 
\end{eqnarray}
where $\phi / \sqrt{2} = \Re[\Phi]$, and 
  we have chosen the renormalization scale to be the VEV of $\Phi$ ($\langle \phi \rangle =M$).  
Here, the coefficient of the one-loop quantum corrections is given by 
\begin{eqnarray}
\beta_\Phi = \frac{1}{16 \pi^2} 
		\left[ 20\lambda_2^2 
			+ 96 g_{BL}^4 - \sum_i(Y_N^i)^4 \right] 
	 \simeq  \frac{1}{16 \pi^2} 
	    \left[ 96 g_{BL}^4 - \sum_i(Y_N^i)^4 \right] , 
\end{eqnarray}
 where in the last expression, we have used $\lambda_2^2 \ll  g_{BL}^4$  as usual in the CW mechanism. 
The stationary condition $\left. dV/d\phi\right|_{\phi=M} = 0$ leads to 
\begin{eqnarray}
   \lambda_2 = \frac{11}{6} \beta_\Phi, 
\label{eq:stationary}
\end{eqnarray} 
 and this $\lambda_2$ is nothing but the renormalized quartic coupling at $M$ defined as 
\begin{eqnarray}
 \lambda_2 = \frac{1}{3 !}\left. \frac{d^4V(\phi)}{d \phi^4} \right|_{\phi=M}. 
\end{eqnarray}  
For more detailed discussion, see \cite{Khoze}.

Associated with this radiative U(1)$_{B-L}$ symmetry breaking,  
  the $Z^\prime$ boson and the right-handed Majorana neutrinos acquire their masses as 
\begin{eqnarray}
  M_{Z^\prime}  = 2 g_{BL} M, \;  \;  \; M_N^i = \frac{Y_N^i}{\sqrt{2}} M. 
\end{eqnarray} 
In this paper, we assume degenerate masses for the three Majorana neutrinos,  
 $Y_N^i = y_N$ (equivalently, $M_N^i=M_N$) for all $i=1,2,3$, for simplicity. 
The U(1)$_{B-L}$ Higgs boson mass is given by 
\begin{eqnarray}
  M_\phi^2 = \left. \frac{d^2 V}{d\phi^2}\right|_{\phi=M}  
                 =\beta_\Phi M^2  
   \simeq  \frac{3}{8 \pi^2}  \frac{ M_{Z^\prime}^4 - 2 M_N^4}{M^2}. 
\label{Eq:mass_phi}
\end{eqnarray} 
When the Majorana Yukawa coupling is negligibly small, this reduces to the well-known relation 
  derived in the radiative symmetry breaking by the CW mechanism \cite{CW}. 
For a sizable Majorana mass, this formula indicates that 
  the potential minimum disappears for $M_N >  M_{Z^\prime}/2^{1/4}$, 
  leading to the upper bound on the right-handed neutrino mass 
  in order for the U(1)$_{B-L}$ symmetry to be broken radiatively.

Once the U(1)$_{B-L}$ gauge symmetry is radiatively broken by the CW mechanism, 
  the electroweak symmetry is subsequently triggered through the coupling $\lambda_3$.  
With $\langle \phi \rangle =M$, the SM Higgs potential is given by  
\begin{equation}
  V(h) = \frac{\lambda}{4}h^4 + \frac{\lambda_{3}}{4} M ^2 h^2,  
\end{equation}
where $H = 1/\sqrt{2}\, (0 \; \; h)^T$ in the unitary gauge.
Choosing $\lambda_3 < 0$, the electroweak symmetry is broken in the same way as in the SM \cite{IOO1}. 
However, the crucial difference from the SM is that in our model the electroweak symmetry breaking 
   originates from the radiative breaking of the U(1)$_{B-L}$ gauge symmetry. 
At the tree level, the stationary condition $V^\prime |_{h=v} = 0$ leads to the relation 
   $|\lambda_3|= 2 \lambda (v/M)^2$, and the Higgs boson mass $m_h$ is given by 
\begin{equation}
  m_h^2 = \left. \frac{d^2 V}{dh^2} \right|_{h=v} = |\lambda_3| M^2 = 2 \lambda v^2. 
\label{Eq:mass_h}
\end{equation}
In the following renormalization group analysis, this relation, 
  $\lambda_3=- m_h^2/M^2$, is used as the boundary condition for $\lambda_3$ at the normalization scale $\mu=M$. 
Since $M \gtrsim 3$ TeV by the LEP constraint \cite{LEPupdate, LEP2A, LEP2B},  $|\lambda_3| \lesssim 10^{-3}$. 
With such a small $\lambda_3$, the back reaction to the $B-L$ Higgs sector through $\lambda_3 v^2$ 
  is negligibly small, and this fact allows us to treat the two Higgs sectors separately.\footnote{
As discussed in Ref.~\cite{IOO2}, this very small $|\lambda_3|$, through which the $B-L$ Higgs can mix with the SM Higgs, 
  makes the experimental search for the $B-L$ Higgs boson very challenging. 
}

\section{Electroweak vacuum stability} 
In the context of the classically conformal U(1)$_{B-L}$ extended model discussed in the previous sections, 
   we now investigate a possibility to solve the electroweak vacuum instability problem. 
The electroweak vacuum stability has been investigated in the minimal $B-L$ model~\cite{Coriano} (see also \cite{CKL}), 
   and the parameter regions for which the electroweak vacuum is stable have been identified. 
A crucial difference in our analysis from the previous one is that 
   our model is classically conformal and the gauge symmetry breaking originates from the CW mechanism. 
Hence, we have constraints on the initial values  of $\lambda_2$ and $\lambda_3$ at the scale $M$, 
   and it is nontrivial to solve the electroweak vacuum instability problem. 
In the classically conformal extension of the SM 
   the electroweak vacuum stability has been investigated though the renormalization group analysis 
   at the one loop level in \cite{Khoze, OOT},  
   it turns out that there is no parameter region to keep the electroweak vacuum stable. 
In the following, we extend the renormalization group analysis to the two-loop level,
   and examine if the vacuum instability can be resolved by the higher order corrections.

In our analysis, we employ the SM renormalization group (RG) equations at the two-loop level~\cite{RGEs}  
  from the top quark pole mass to the U(1)$_{B-L}$ Higgs VEV ($M$), and connect the RG equations  
  to those of the minimal U(1)$_{B-L}$ extended SM at the two-loop level.\footnote{
To generate the RG equations at the two-loop level for the minimal U(1)$_{B-L}$ model, 
  we have used SARAH~\cite{SARAH}.  
For a complete RG analysis at the two-loop level, we need to take into account the threshold corrections 
  at the 1-loop level to match the 2-loop RG evolutions at $M$. 
The most important corrections is to top Yukawa coupling at $M$ 
  since the electroweak vacuum instability problem is very sensitive to the input of top Yukawa coupling. 
We have estimated the threshold corrections to be of the order of $y_t \times (1/3)^2 \alpha_{BL}/(4 \pi)$ 
  through the $Z^\prime$ boson loop diagrams,  
  which changes the top Yukawa input at $M$ by ${\cal O}(0.01\%)$ for $\alpha_{BL}=0.012$ (see Fig.~\ref{fig4}), 
  or equivalently  ${\cal O}(0.01 {\rm GeV})$ in terms of top quark mass. 
Since we have neglected the threshold corrections in our analysis, 
  our results in this paper have a theoretical uncertainty of O(0.01 GeV) in the top quark mass. 
As can be seen from Fig.~\ref{fig3}, the uncertainty at this size is negligibly small. 
}
All formulas used in our analysis are listed in Appendices. 
As is well-known, the RG evolutions of the Higgs quartic coupling is sensitive to 
  the input values of the Higgs boson and top quark masses. 
For inputs for the Higgs boson mass and top quark pole mass, 
   we adopt the result from the combined analysis by the ATLAS and the CMS experiments 
   for the Higgs boson mass measurement in the range of 
    $m_{h}=$125.09$\pm$ 0.21 (stat.)$\pm$ 0.11 (syst.) GeV~\cite{Higgs1} and 
   the recent result of top quark mass measurement by the CMS experiments~\cite{top1}
   in the range of $m_{t}=$ 172.38$\pm$ 0.10$\pm$ 0.65.

The RG evolutions of the Higgs quartic coupling are shown in Fig.~\ref{fig1} for two different values 
   of $m_h=125.09$  GeV (left panel) and $125.41$ GeV (right panel) with a fixed $m_t=171.63$ GeV. 
Here we have fixed the other parameters as $g_{BL}=0.314$,  $g_{YB}=g_{BY}=0$ and $y_N=0$ at $\mu=M=4$ TeV. 
The solid lines denote the RG evolutions of the Higgs quartic coupling in our model, 
  while the dashed lines denote those in the SM. 
We can see that in our model, the Higgs quartic coupling remains positive up to the Planck scale, 
  $M_{Pl}=1.2 \times 10^{19}$ GeV, and therefore the electroweak vacuum becomes stable. 
As the same as in the SM \cite{RGEs}, the situation becomes better with an increasing (decreasing) value 
  of $m_h$ ($m_t$) for a fixed value of the $m_t$ ($m_h$).  

\begin{figure}
\begin{center}
\includegraphics[scale=0.725]{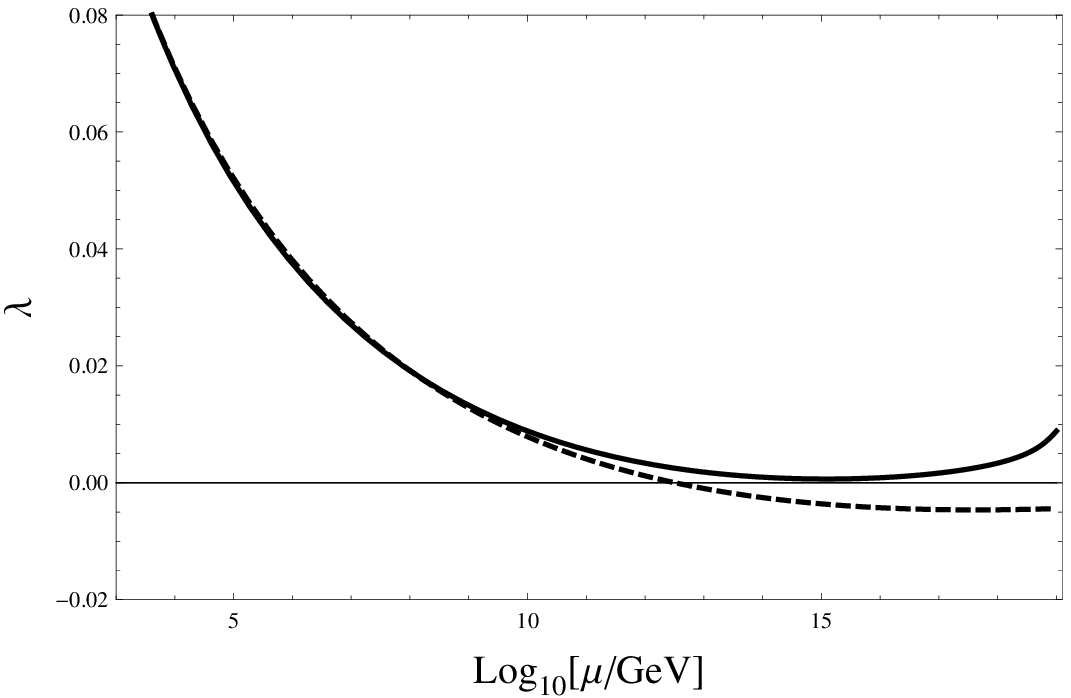}  \; 
\includegraphics[scale=0.71]{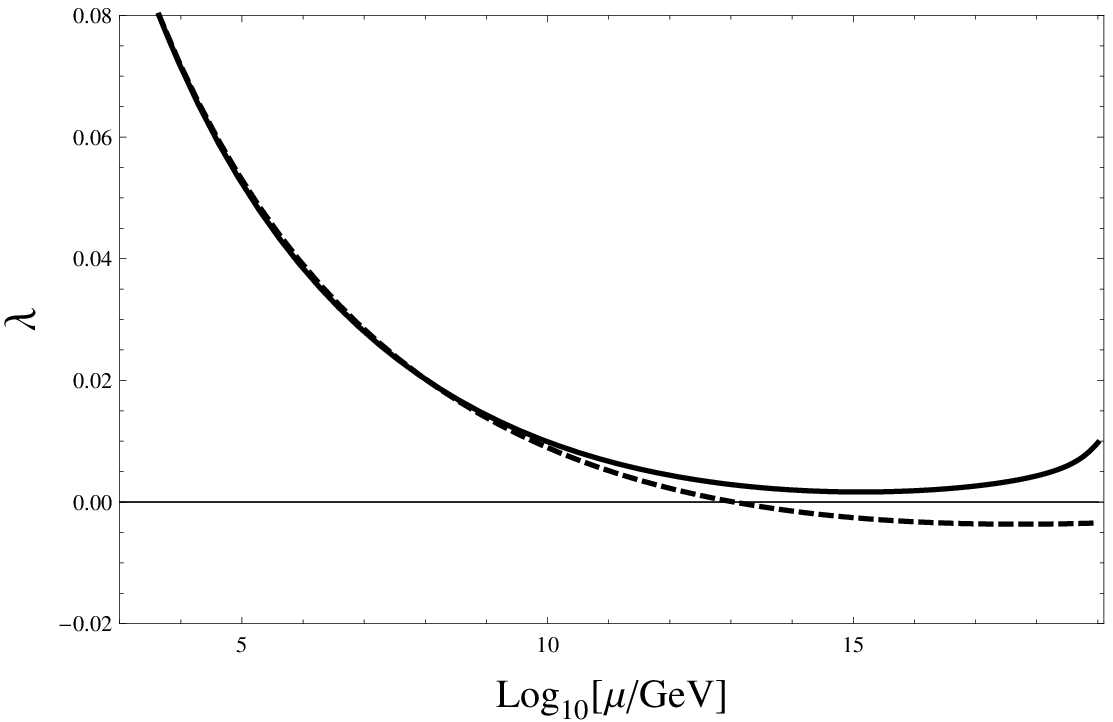}
\end{center}
\caption{
The renormalization group evolution of the Higgs quartic coupling ($\lambda$) 
    in the $B-L$ model (solid lines), along with the one in the SM (dashed lines).  
 We have taken $m_h=125.09$ GeV (left panel) and $m_h=125.41$ GeV (right panel) 
    with the fixed values  of $M= 4.0$ TeV and $m_t=$ 171.63 GeV.    }
\label{fig1}
\end{figure}

\begin{figure}
\begin{center}
\includegraphics[scale=0.84]{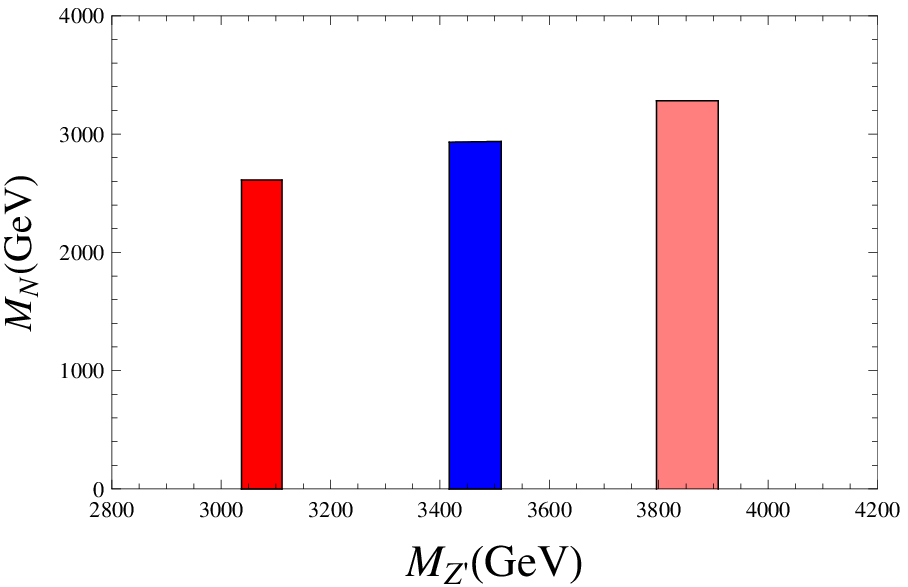}  \;
\includegraphics[scale=0.84]{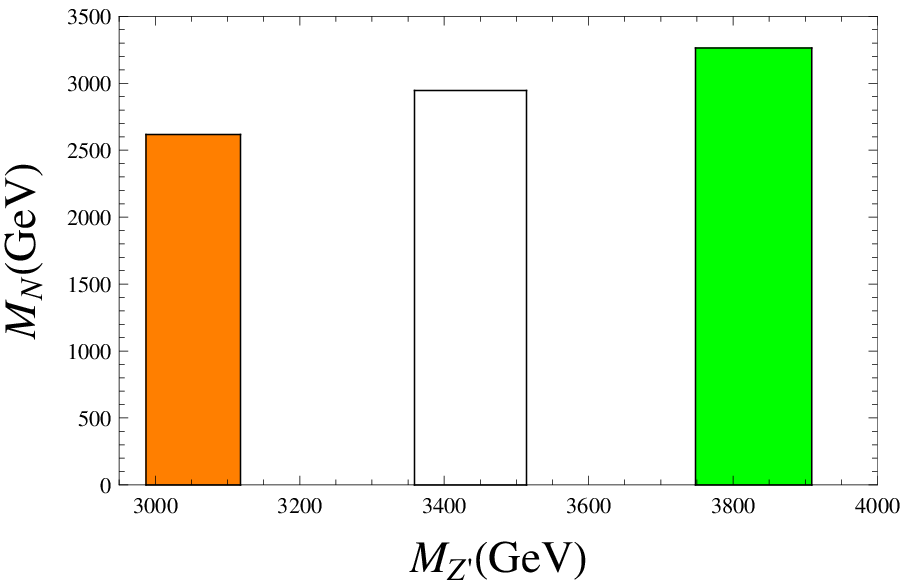}\\
\includegraphics[scale=0.84]{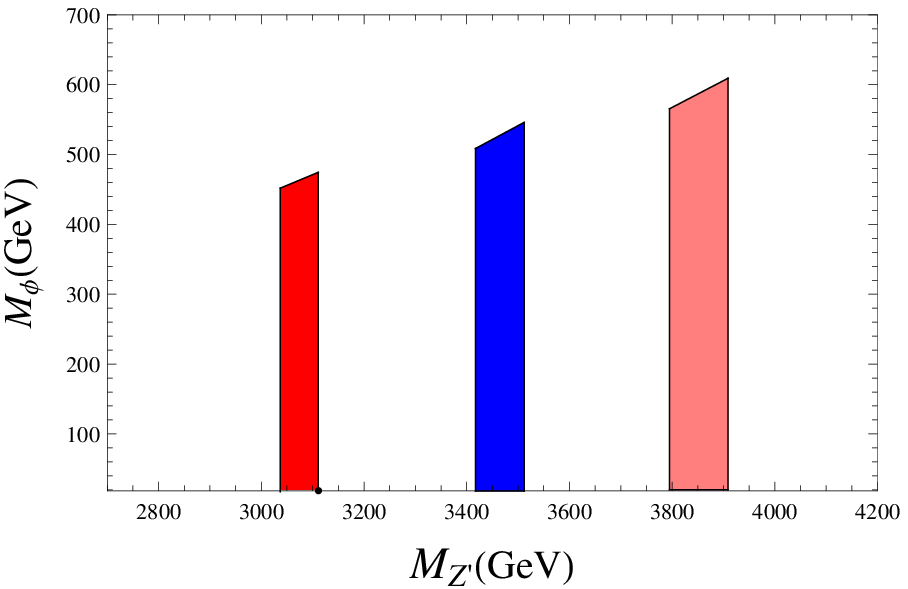}  \; 
\includegraphics[scale=0.84]{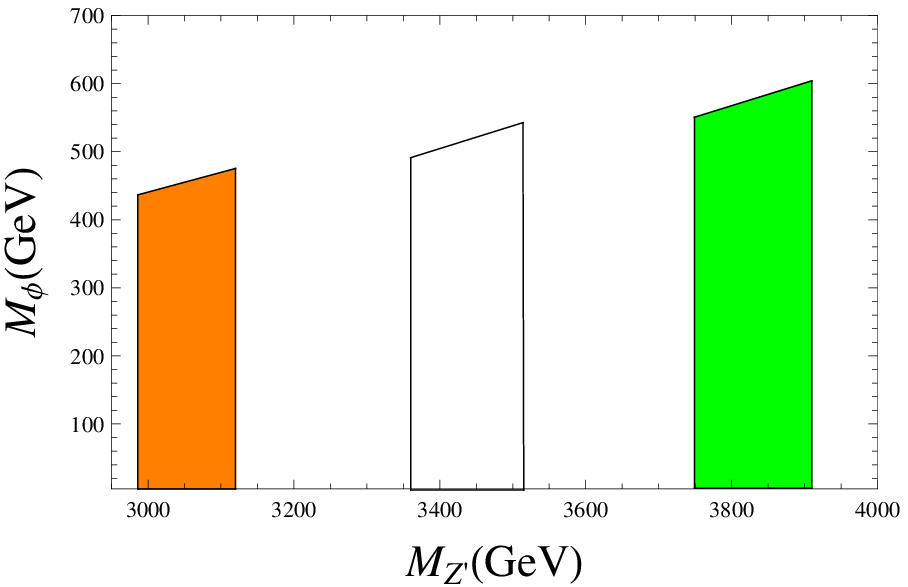}\\
\end{center}
\caption{
The results of parameter scans for $M_{Z^\prime}$ and $M_N$.  
We have used  $m_h=125.09$ GeV (two panels in the left column) and $m_h=125.41$ GeV (two panels in the right column), 
   for the fixed value of $m_t=$ 171.63 GeV.   
In each panel,  three regions from left to right correspond to $M=3.5$, $4.0$ and $4.5$ TeV, respectively. 
}
\label{fig2}
\end{figure}

In order to identify parameter regions to keep the electroweak vacuum stable, 
   we perform parameter scans for the free parameters $M_{Z^\prime}$ and $M_N$ 
   with fixed values of $M=3.5$, $4.0$ and $4.5$ TeV. 
Here, we have used the same values for $m_h$ and $m_t$ as in Fig.~\ref{fig1}. 
In this analysis, we impose the following conditions for the running couplings at $M \leq \mu \leq M_{Pl}$: 
  the stability of the Higgs potential  ($\lambda,  \lambda_2 > 0$ and $|\lambda_3|^2 < 4 \lambda \lambda_2$), 
  and the conditions that all the running couplings remain in the perturbative regime, namely, 
  $g_i^2 (i=1,2,3), g_{BL}^2, g_{YB}^2,  g_{BY}^2 <4 \pi$ and $\lambda, \lambda_{2,3} < 4 \pi $. 
The results are shown in Fig.~{\ref{fig2}}. 
In this Figure, we also show the $B-L$ Higgs boson mass by using Eq.~(\ref{Eq:mass_phi}). 
As we expect, the allowed region becomes larger as $m_h$ is increased.

We also perform parameter scan for various values of $m_h$ and $m_t$ in the ranges 
  of $124.68 \leq m_h/{\rm GeV} \leq 125.32$ and $171.63 \leq m_t/{\rm GeV} \leq 173.13$, 
  with fixed values of $M=2$ and $4$ TeV.  
The results are shown in Fig.~\ref{fig3} for $M=2$ TeV (left panel) and $M=4$ TeV (right panel). 
The parameter sets inside of the triangles satisfy all constraints of the electroweak vacuum stability 
  and the perturbativity of the running couplings.  
For a fixed $m_h$, there is an upper bound on  $m_t$, 
  or equivalently, there is a lower bound on $m_h$ for a fixed $m_t$. 
The allowed region for $M=4$ TeV is more restricted than the one for $M=2$ TeV.  
When we increase the $M$ value further, the allowed region disappears (see Fig.~\ref{fig4}).

\begin{figure}
\begin{center}
\includegraphics[scale=0.875]{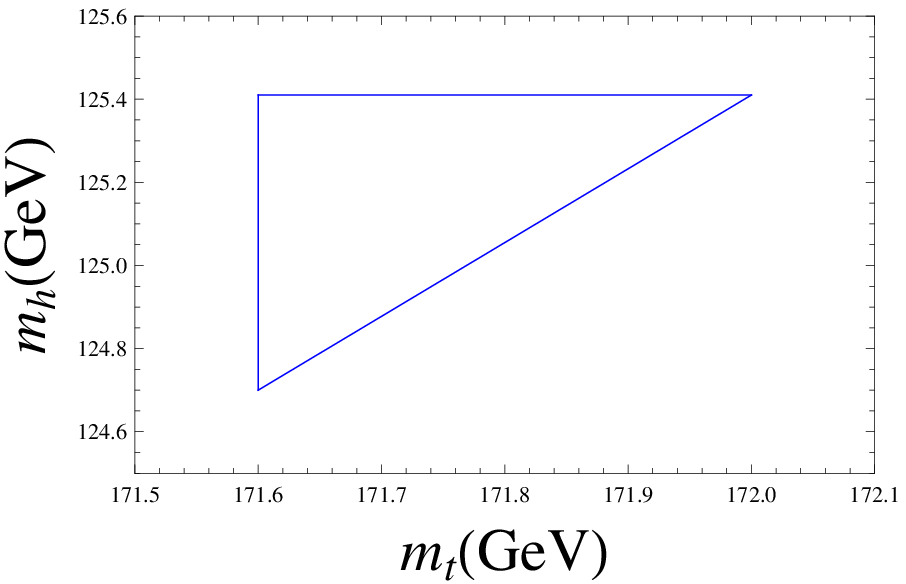}  \; 
\includegraphics[scale=0.875]{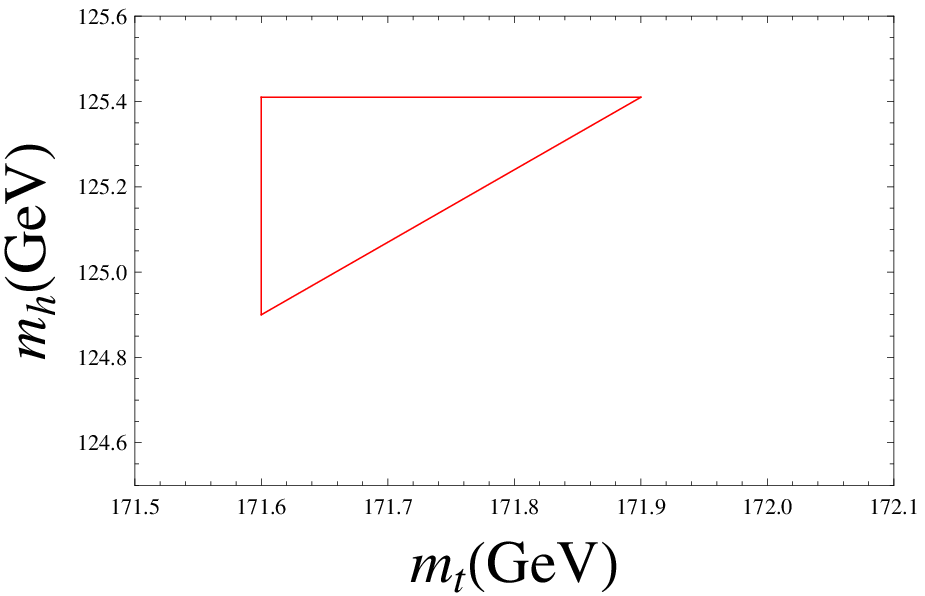}
\end{center}
\caption{
The results of parameter scans for various values of $m_h$ and $m_t$ in the ranges 
  of $124.68 \leq m_h/{\rm GeV} \leq 125.32$ and $171.63 \leq m_t/{\rm GeV} \leq 173.13$, 
  with the fixed values of $M=2$ TeV (left panel) and $M=4$ TeV (right panel). 
}
\label{fig3}
\end{figure}

\begin{figure}
\begin{center}
\includegraphics[scale=0.9]{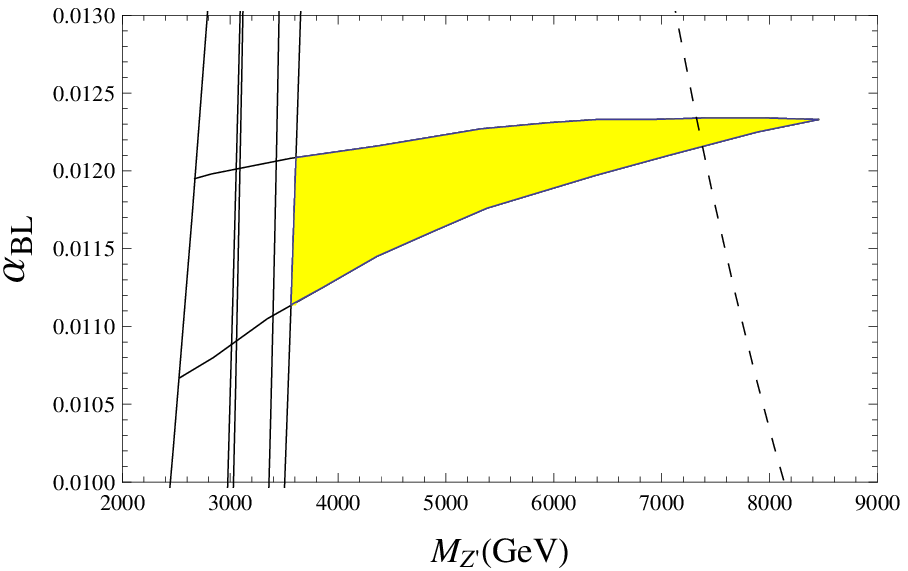} 
\includegraphics[scale=0.9]{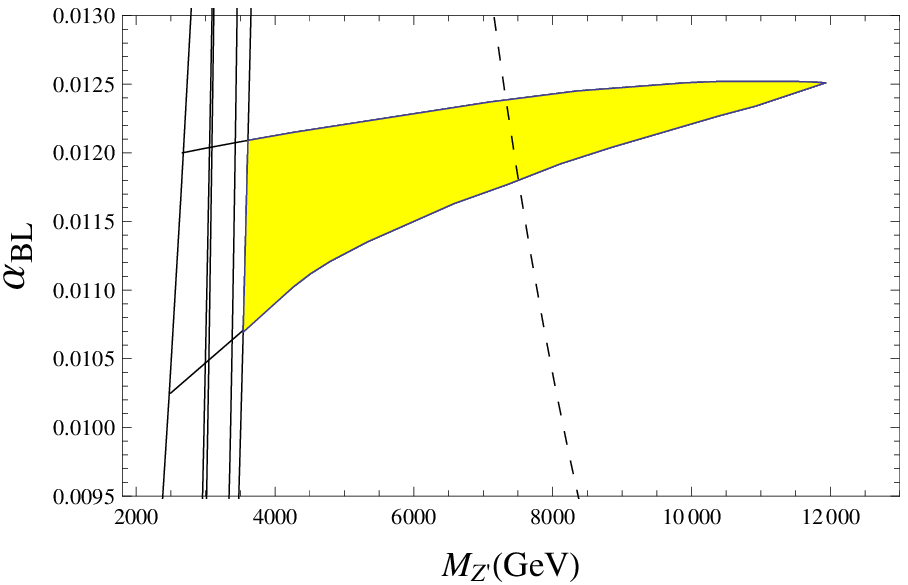} 
\end{center}
\caption{
The results of parameter scans for various values of $\alpha_{BL}=g_{BL}^2/(4 \pi)$ and $M_{Z^\prime}$. 
We have used $m_h = 124.77$ GeV (left panel) and $125.09$ GeV (right panel) 
 with $m_t = 171.63$ GeV. 
The regions inside the shaded triangles satisfy all the constraints. 
The vertical solid lines from left to right correspond to the limits from the LEP, 
   the ATLAS with the LHC Run-1, the CMS with the LHC Run-1,
   the CMS with the LHC Run-2 and the ATLAS with the LHC Run-2, respectively. 
The naturalness argument prefers the regions on the left sides of the diagonal dashed lines. 
}
\label{fig4}
\end{figure}

Finally, we show in Fig.~\ref{fig4} the results of our parameter scans for various values of $g_{BL}$ and $M$, 
  with $m_h=124.77$ GeV (left panel) and 125.09 GeV (right panel) for $m_t=171.63$ GeV. 
In this Figure, we present the results with $\alpha_{BL}=g_{BL}^2/(4 \pi)$ 
   and $M_{Z^\prime}$ by using the mass formula $M_{Z^\prime}=2 g_{BL} M$. 
Here we have considered not only the conditions of the electroweak vacuum stability and the perturbativity, 
   but also the current collider bounds.  
The search for effective 4-Fermi interactions mediated by the $Z^\prime_{BL}$ boson 
   at the LEP leads to a bound~\cite{LEPupdate} (see also \cite{LEP2A, LEP2B})
\bea
   \frac{M_{Z^\prime}}{g_{BL}} \geq 6.9 \; {\rm TeV} 
\eea 
at 95\% confidence level.  
The ATLAS and the CMS collaborations have searched for $Z^\prime$ boson resonance 
  at the LHC Run-1 with $\sqrt{s}=8$ TeV. 
The most stringent bounds on the $Z^\prime$ boson production cross section times branching ratio 
  have been obtained by using the dilepton final state. 
For the so-called sequential SM $Z^\prime$ model~\cite{ZpSSM}, where the $Z^\prime$ boson has exactly 
  the same couplings with the SM fermions as those of the SM $Z$ boson, 
  the cross section bounds lead to lower bounds on the $Z^\prime$ boson mass
  as $M_{Z^\prime} \geq 2.90$ TeV from the ATALS analysis~\cite{ATLAS8} and 
  $M_{Z^\prime} \geq 2.96$ TeV from the CMS analysis~\cite{CMS8}, respectively.  
Very recently, these bounds have been updated by the ATLAS~\cite{ATLAS13} and CMS~\cite{CMS13} analysis 
   with the LHC Run-2 at $\sqrt{s}=13$ TeV as 
  $M_{Z^\prime} \geq 3.4$ TeV (ATLAS) and $M_{Z^\prime} \geq 3.15$ TeV (CMS), respectively.     
We interpret theses ATLAS and CMS results to the $B-L$ $Z^\prime$ boson case.  
In our model, the U(1)$_{B-L}$ gauge coupling is a free parameter, and for a fixed gauge coupling 
   we can read off the lower limit on the $Z^\prime$ boson mass from the ATLAS and CMS cross section bounds. 
In this way, we can find an upper (lower) bound on the the U(1)$_{B-L}$ gauge coupling $\alpha_{BL}=g_{BL}^2/(4 \pi)$ 
   ($Z^\prime$ boson mass $M_{Z^\prime}$) 
    as a function of $M_{Z^\prime}$ ($\alpha_{BL}$).  
In interpreting the ATLAS and the CMS results to the $B-L$ model,  
   we follow a strategy presented in detail in \cite{NOSO} (see also \cite{LHCZp}).   
In Fig.~\ref{fig4}, the vertical solid lines correspond to the bounds from the LEP result,  
   the ATLAS with the LHC Run-1, the CSM with the LHC Run 1, 
   the CMS with the LHC Run-2 and the ATLAS with the LHC Run-2, from left to right.  
The parameters inside the shaded triangles satisfy all the constraints. 
Naturalness bound, which will be obtained in the next section, is also shown as the dashed lines.  
In, for example, Ref.~\cite{Basso}, the search reach of the $Z^\prime$ boson 
   at the LHC Run 2 with a 14 TeV collider energy and a 100/fb luminosity 
   is obtained as $M_{Z^\prime} \simeq 5$ TeV for $\alpha_{BL}\simeq 0.01$.  
A large potion of the allowed regions presented in Fig.~\ref{fig4} can be tested in the near future. 
The (indirect) search reach of the future $e^+ e^-$ linear collider 
   with a 1 TeV collider energy can be as large as 10 TeV (see, for example, \cite{IOO2}), 
   and almost of all allowed regions presented in Fig.~\ref{fig4} can be covered.

\section{Constraints from Naturalness}
Once the U(1)$_{B-L}$ gauge symmetry is radiatively broken by the CW mechanism, 
    the masses for the $Z^\prime$ boson and the Majorana neutrinos are generated, 
    which in general create self-energy corrections to the SM Higgs doublet.  
If the $B-L$ gauge symmetry breaking scale is very large, 
    the self-energy corrections may exceed the electroweak scale and 
    require us to fine-tune the model parameters in reproducing the correct electroweak scale. 
Two major corrections have been discussed in \cite{IOO1, IOO2}: 
one is one-loop corrections with the Majorana neutrinos, and the other is 
    two-loop corrections involving the $Z^\prime$ boson and the top quark. 
In the calculations of the self-energy corrections in \cite{IOO2}, 
   the cutoff procedure with  the Planck scale cutoff is applied to derive the naturalness bounds. 
Although this treatment is good for rough estimates, 
   in order to derive more accurate naturalness bounds 
   we will renormalize the loop corrections properly in this section.

Since the original theory is classically conformal and defined as a massless theory,  
   the self-energy corrections to the SM Higgs doublet originates 
   from corrections to the quartic coupling $\lambda_3$.  
Thus, what we calculate to derive the naturalness bounds is quantum corrections to 
   the term $\lambda_3 h^2 \phi^2$ in the effective Higgs potential.  
For the one-loop diagram involving the Majorana neutrinos (for the Feynman diagram, see Fig.~3 in \cite{IOO2}), 
   we calculate the effective potential as 
\bea 
\Delta V_{\rm 1-loop} \supset - \frac{|Y_D|^2 |Y_N|^2} {16 \pi^2} h^2  \phi^2 \left( \ln [\phi^2] + C \right), 
\eea 
where the logarithmic divergence and the terms independent of $\phi$ are all encoded in $C$.  
By adding a counter term, we renormalize the coupling $\lambda_3$ with the renormalization condition, 
\bea 
   \frac{\partial^4}{\partial h^2 \partial \phi^2} V_{\rm eff} \Big|_{h=0, \phi=M} = \lambda_3, 
\eea  
   where $V_{\rm eff}$ is the sum of the tree-level potential and $\Delta V_{\rm 1-loop}$, 
   and $\lambda_3$ is the renormalized coupling. 
As a result, we obtain 
\bea 
  V_{\rm eff} \supset \left[  
    \frac{1}{4} \lambda_3  - \frac{|Y_D|^2 |Y_N|^2} {16 \pi^2}  \left( \ln\left[\frac{\phi^2}{M^2}\right] -3 \right) 
     \right] h^2 \phi^2.  
\eea
Substituting $\phi=M$, we obtain the SM Higgs self-energy correction as 
\bea    
   \Delta m_h^2 =  \frac{3 |Y_D|^2 |Y_N|^2} {8 \pi^2}  M^2 \sim \frac{3 m_\nu M_N^3}{4 \pi^2 v^2}
 \label{Eq:Ncontribution}
\eea
where we have used the seesaw formula, $m_\nu \sim Y_D^2 v^2/M_N$ \cite{seesaw}. 
If $\Delta m_h^2$ is much larger than the electroweak scale,  
  we need a fine-tuning of the tree-level Higgs mass ($|\lambda_3| M^2/2$) 
  to reproduce the correct Higgs VEV, $v=246$ GeV. 
Here, we introduce the naturalness condition as 
\begin{eqnarray}
  \delta  = \frac{m_h^2}{2 |\Delta m_h^2|} \gtrsim 1.   
  \label{Eqn:naturalness}
\end{eqnarray}
For example, when the light neutrino mass scale is around $m_\nu \simeq 0.1$ eV after the seesaw mechanism, 
   we have an upper bound for the  Majorana mass as $M_N \lesssim 4 \times 10^6$ GeV.

For the two-loop diagrams involving $Z^\prime$ boson and top quark 
   (for the Feynman diagrams, see Fig.~4 in \cite{IOO2}), we have 
\bea 
 \Delta V_{\rm 2-loop} \supset - \frac{2 \alpha_{BL}^2 m_t^2} {\pi^2 v^2} h^2  \phi^2 \left( \ln [\phi^2] + C \right), 
\eea 
where the logarithmic divergence and the terms independent of $\phi$ are all encoded in $C$.  
Following the same strategy as the above, we obtain 
\bea    
   \Delta m_h^2 =  \frac{3 \alpha_{BL} m_t^2} {4 \pi^2 v^2}  M_{Z^\prime}^2. 
 \label{Eq:Zpcontribution}
\eea
The dashed lines shown in Fig.~\ref{fig4} are plotted by using the condition $\delta=1$ in Eq.~(\ref{Eqn:naturalness}).

\section{Conclusions}
We have considered the minimal $B-L$  extension of the Standard Model,  
   where the anomaly-free global $B-L$ symmetry in the Standard Model is gauged 
   and three right-hand neutrinos and a $B-L$ Higgs field are introduced. 
This model is very simple and well-motivated, since the right-handed neutrinos 
  acquire their Majorana masses associated with the $B-L$ gauge symmetry breaking,  
  and the seesaw mechanism for the neutrino mass generation is automatically implemented. 
Motivated by the argument that the Higgs model can be free from the the gauge hierarchy problem 
   once the classically conformal symmetry is imposed in the model, 
   we have introduced the classically conformal symmetry to the minimal $B-L$ model. 
In this context, the $B-L$ symmetry is radiatively broken by the Coleman-Weinberg mechanism 
   and this breaking is the sole origin of all mass parameters in the model. 
The electroweak symmetry breaking is realized by the negative mass term for the Higgs doublet, 
   which is subsequently generated through the $B-L$ gauge symmetry breaking. 
Therefore, the electroweak symmetry breaking originates from the radiative $B-L$ gauge symmetry breaking.

In the context of the classically conformal $B-L$ model, we have investigated 
   the electroweak vacuum instability problem. 
With the measured Higgs boson mass around $125$ GeV,  it turns out that 
   the electroweak vacuum is not the true minimum in the the effective Higgs potential 
   of the Standard Model.  
In other words, the electroweak symmetry is radiatively broken at some energy 
   much higher than the electroweak scale.    
This ruins the theoretical consistency of our model that the radiative $B-L$ symmetry breaking 
   is the sole origin of the mass. 
We have analyzed the renormalization group evolutions of the model couplings 
   at the two-loop level with the recent results of the Higgs boson mass and top quark mass 
   measurements at the LHC.  
We have identified parameter regions which satisfy the conditions of 
   the stability of the electroweak vacuum and the perturbativity of the running couplings, 
   as well as the current collider bounds from the search for the $B-L$ gauge boson, 
   in particular, at the LHC Run-2.

In addition, we have considered the naturalness of the electroweak scale  
   against self-energy corrections for the Higgs doublet.  
We have refined the previously obtained results in a theoretically consistent way 
   for the Coleman-Weinberg effective potential, and derived the naturalness bounds 
   on the $B-L$ gauge boson and the right-handed neutrino masses. 
The allowed regions satisfying the naturalness bounds can be tested 
   in the future collider experiments.

\section*{Acknowledgements}
This work is supported in part by the United States Department of Energy Grant, No. DE-SC0013680. 

\appendix
\section{The beta functions for the SM couplings}
\subsection{The one-loop beta functions for the SM gauge couplings}
\bea
\beta_{g_{1}}^{(1)} = \frac{41}{10} g_{1}^3, \; \;
\beta_{g_{2}}^{(1)} = -\frac{19}{6} g_{2}^3,  \; \; 
\beta_{g_{3}}^{(1)} = -7 g_{3}^3. 
\eea
\subsection{The one-loop beta function for the top Yukawa coupling}
\bea
\beta_{y_{t}}^{(1)} &=& y_{t}\bigg(-\frac{17}{20} g_{1}^2-\frac{9 g_{2}^2}{4}- 8 g_{3}^2+\frac{9 y_{t}^2}{2}\bigg).
\eea
\subsection{The one-loop beta function for the quartic Higgs coupling}
\bea
\beta_{\lambda}^{(1)} = -\lambda \bigg(\frac{9 g_{1}^2}{5}+9 g_{2}^2\bigg)+\frac{9}{4} \bigg(\frac{2}{5} g_{1}^2 g_{2}^2+ 
        \frac{3 g_{1}^4}{25}+  g_{2}^{4}\bigg)+12 \lambda^{2}+12 \lambda y_{t}^{2}-12 y_{t}^{4}. 
\eea

\subsection{The two-loop beta functions for the gauge couplings}
\bea
\beta_{g_{1}}^{(2)} &=& g_{1}^3 \Big(\frac{199 g_{1}^2}{50}+ \frac{27 g_{2}^2}{10}+ \frac{44 g_{3}^2}{5}- \frac{17 y_{t}^2}{10}\Big),
\nonumber \\
\beta_{g_{2}}^{(2)} &=& g_{2}^3 \Big(\frac{9 g_{1}^2}{10}+ \frac{35 g_{2}^2}{6}+ 12 g_{3}^2- \frac{3 y_{t}^2}{2}\Big),
\nonumber \\
\beta_{g_{3}}^{(2)} &=& g_{3}^3 \Big(\frac{11 g_{1}^2}{10}+ \frac{9 g_{2}^2}{2}-26 g_{3}^{2}- 2 y_{t}^2\Big).  
\eea

\subsection{The two-loop beta function for the top Yukawa coupling}
\bea
\beta_{y_{t}}^{(2)} &=& y_{t}\bigg(y_{t}^2 \bigg(\frac{393 g_{1}^2}{80}+\frac{225 g_{2}^2}{16}+36 g_{3}^2\bigg)- 
                     \frac{9}{20} g_{1}^2 g_{2}^2+ \frac{19}{15} g_{1}^2 g_{3}^2+\frac{1187 g_{1}^4}{600} \nonumber \\
                    &+& 9 g_{2}^2 g_{3}^2- \frac{23 g_{2}^{4}}{4}-108 g_{3}^4 + 
                    \frac{3 \lambda ^2}{2}- 6\lambda  y_{t}^2- 12 y_{t}^{4}\bigg). 
\eea

\subsection{The two-loop beta function for the Higgs quartic coupling}
\bea
\beta_{\lambda}^{(2)} &=& \bigg( 10 \lambda  y_{t}^2 \bigg(\frac{17 g_{1}^2}{20}+\frac{9 g_{2}^2}{4}+8 g_{3}^2\bigg)
                      + 18 \lambda ^2 \bigg(\frac{3 g_{1}^2}{5}+3g_{2}^2\bigg) \nonumber \\
                      &-& \lambda  \bigg(-\frac{117}{20} g_{1}^2 g_{2}^2-\frac{1887}{200} g_{1}^4+\frac{73g_{2}^4}{8}\bigg)
                       -\frac{1677}{200} g_{1}^4 g_{2}^2- \frac{289}{40} g_{1}^2 g_{2}^4 \nonumber \\
                      &-&  \frac{3}{5} g_{1}^2 y_{t}^2 \bigg(\frac{57 g_{1}^2}{10}-21 g_{2}^2\bigg)-\frac{3411 g_{1}^6}{1000}
                      - \frac{16}{5} g_{1}^2 y_{t}^4+\frac{305 g_{2}^6}{8} \nonumber \\
                      &-& \frac{9}{2} g_{2}^4 y_{t}^2-64 g_{3}^2 y_{t}^4-78 \lambda ^3-72 \lambda ^2 y_{t}^2
                       - 3\lambda  y_{t}^4+60 y_{t}^6\bigg). 
\eea

In our analysis, we numerically solve the SM RG equations with the following boundary conditions at $\mu=m_t$ 
   \cite{RGEs}\footnote{
We have employed the boundary conditions in arXiv:1307.3536v4.
}:   
\begin{eqnarray}
g_3(m_t) &=& \sqrt{\frac{5}{3}} \left[
              1.1666 + 0.00314 \left( \frac{\alpha_3(m_Z) - 0.1184}{0.0007} \right) 
		    -  0.00046 \left( \frac{m_t}{\rm GeV} - 173.34 \right) \right], \nonumber \\
g_2(m_t) &=& 0.64779 + 0.00004 \left( \frac{m_t}{\rm GeV} - 173.34 \right) 
			+ 0.00011 \left( \frac{m_W - 80.384{\rm GeV}}{0.014{\rm GeV}} \right), \nonumber \\
g_1(m_t) &=& 0.35830 + 0.00011 \left( \frac{m_t}{\rm GeV} - 173.34 \right) 
			- 0.00020\left(  \frac{m_W - 80.384{\rm GeV}}{0.014{\rm GeV}} \right),\nonumber \\
y_t(m_t) &=& 0.93690 + 0.00556 \left( \frac{m_t}{\rm GeV} - 173.34 \right) 
			- 0.00042 \left( \frac{\alpha_3(m_Z) - 0.1184}{0.0007} \right), \nonumber \\
\lambda(m_t) &=& 0.12604 + 0.00206 \left( \frac{m_h}{\rm GeV} - 125.15 \right) 
			- 0.00004 \left( \frac{m_t}{\rm GeV} - 173.34\right).  
\label{Eq:SM_BC_lambda_H}
\end{eqnarray}
We have used the inputs, $\alpha_3(m_Z) = 0.1184$ and $m_W=80.384$ GeV.


\section{The beta functions for the couplings in the U(1)$_{B-L}$ extended SM}
\subsection{The one-loop beta functions for the gauge couplings}
\bea
\beta_{g_{1}}^{(1)} &=& \frac{1}{10} \left( g_{1} \Big(32 \sqrt{\frac{5}{3}} g_{BL} g_{BY}+41 g_{BY}^2+180 g_{YB}^2  \right) 
                 \nonumber \\
                  &+& 32 \sqrt{10} g_{1}^2 g_{YB}+41 g_{1}^3+4 g_{BY} g_{YB} \Big(15 \sqrt{6} g_{BL}+ 4 \sqrt{10} g_{BY}\Big)\Big), 
\nonumber \\
\beta_{g_{2}}^{(1)} &=&-\frac{19}{6} g_{2}^3, \nonumber \\
\beta_{g_{3}}^{(1)} &=& -7 g_{3}^3, \nonumber \\
\beta_{g_{BY}}^{(1)} &=& \frac{1}{10} \Big(4 g_{1} g_{YB} \Big(15 \sqrt{6} g_{BL}+4 \sqrt{10} g_{BY} \Big)
           + g_{1}^2 \Big(32 \sqrt{\frac{5}{3}} g_{BL}+41 g_{BY}\Big) \nonumber \\
           &+& g_{BY} \Big(64 \sqrt{\frac{5}{3}} g_{BL} g_{BY}+120 g_{BL}^2+41 g_{BY}^2\Big)\Big), \nonumber \\
\beta_{g_{YB}}^{(1)} &=& \frac{1}{10} \Big( g_{1} \Big(41 \sqrt{\frac{2}{3}} g_{BL} g_{BY} 
           +16 \sqrt{10} \Big(\frac{2 g_{BL}^2}{3}+2 g_{YB}^2\Big)\Big)  \nonumber \\
           &+& 41 g_{1}^2 g_{YB}+4 g_{YB} \Big(8 \sqrt{\frac{5}{3}} g_{BL} g_{BY}
                   +45 \Big(\frac{2 g_{BL}^2}{3}+ g_{YB}^2\Big)\Big)\Big),  \nonumber \\ 
\beta_{g_{BL}}^{(1)} &=& \sqrt{\frac{3}{2}}\Big(\frac{1}{10} \Big(4 \sqrt{\frac{2}{3}} g_{BL} \Big(g_{YB} \Big(4 \sqrt{10} g_{1} 
     +45 g_{YB}\Big)+30 g_{BL}^2\Big)   \nonumber \\ 
     &+&  g{BY} \Big( g_{YB} \Big(41 g_{1}^2+16 \sqrt{10} g_{YB}\Big) 
        + 32 \sqrt{10} g_{YB}^2\Big)+41 \sqrt{\frac{2}{3}} g_{BL} g_{BY}^2\Big) \Big).  
\eea

\subsection{The one-loop beta function for the top Yukawa coupling}
\bea
\beta_{y_{t}}^{(1)} = \frac{9 y_{t}^3}{2}- y_{t} \Big(\sqrt{\frac{5}{3}} g_{BY} g_{BL}+\frac{2}{3} g_{BL}^2
        + \sqrt{\frac{5}{2}} g_{1} g_{YB}+ \frac{17 g_{1}^2}{20}+\frac{9 g_{2}^2}{4}+8 g_{3}^2+\frac{17 g_{BY}^2}{20}+g_{YB}^2\Big). 
\eea
\subsection{The one-loop beta function for the Majorana Yukawa coupling}
\bea
\beta_{Y_{N}}^{(1)} &=& 2 \Big(10 \Big(\frac{Y_{N}}{2}\Big)^3-9 \frac{Y_{N}}{2}\Big(\frac{2}{3} g_{BL}^2+g_{YB}^2\Big)\Big). 
\eea

\subsection{The one-loop beta function for the scalar quartic couplings}
\bea
\beta_{\lambda}^{(1)} &=& 2\Big( \frac{\lambda}{2} \Big(-\frac{9}{5} g_{1}^2-9 g_{2}^2-\frac{9 g_{BY}^2}{5}+12 y_{t}^2\Big) 
     \nonumber \\
           &+& 24 \Big(\frac{\lambda}{2}\Big)^2+\frac{9}{20} g_{1}^2 g_{2}^2+\frac{27}{100} g_{1}^2 g_{BY}^2  
     \nonumber \\
           &+& \frac{27}{200} g_{1}^4 +\frac{9}{20} g_{2}^2 g_{BY}^2+\frac{9 g_{2}^4}{8}+\frac{27 g_{BY}^4}{200}
                  +  \lambda_{3}^2-6 y_{t}^4 \Big), 
     \nonumber \\
\beta_{\lambda_{2}}^{(1)} &=& 2 \Big(-36 \lambda_{2} \Big(\frac{2}{3} g_{BL}^2+ g_{YB}^2\Big)+ 144 g_{YB}^2 g_{BL}^2 
      \nonumber \\
          &+& 48 g_{BL}^4+108 g_{YB}^4+ 10 \lambda_{2}^2+\lambda_{3}^2+3 \lambda_{2} Y_{N}^2-\frac{3}{2} Y_{N}^4\Big), 
      \nonumber \\
\beta_{\lambda_{3}}^{(1)} &=& -\lambda_3 \Big(24 g_{BL}^2+\frac{9}{10} g_1^2+\frac{9}{2} g_2^2
                 +\frac{9 g_{BY}^2}{10}+36 g_{YB}^2\Big) 
       \nonumber \\
          &+& \frac{36}{5} \sqrt{6} g_1 g_{BY} g_{YB} g_{BL}+\frac{36}{5} g_{BY}^2 g_{BL}^2 + \frac{54}{5} g_1^2 g_{YB}^2 
       \nonumber \\
          &+& 4 \lambda_2^2+6 \lambda \lambda_3+8 \lambda_2 \lambda_3 + \lambda_3 \Big(3 Y_{N}^2+6 y_{t}^2\Big). 
\eea

\subsection{The two-loop beta functions for the gauge couplings}
\bea
\beta_{g_{1}}^{(2)} &=& 
            \frac{1}{100} \Big(398 g_{1}^5+328 \sqrt{10} g_{BY} g_{1}^4+398 g_{BY} g_{1}^4
           +328 \sqrt{10} g_{YB} g_{1}^4+270 g_{2}^2 g_{1}^3 
     \nonumber\\ 
         &+& 880 g_{3}^2 g_{1}^3 + 328 \sqrt{10} g_{BY}^2 g_{1}^3 +1318 g_{BY}^2 g_{1}^3
                  + 920 g_{YB}^2 g_{1}^3-170 y_{t}^2 g_{1}^3 
     \nonumber\\        
         &+&  \frac{1840}{3} g_{BL}{}^2 g_{1}^3+ 328 \sqrt{10} g_{BY} g_{YB} g_{1}^3
                 + 3680 g_{BY} g_{YB} g_{1}^3+656 \sqrt{\frac{5}{3}} g_{BY} g_{BL} g_{1}^3 
     \nonumber\\       
          &+& 328 \sqrt{10} g_{BY}^3 g_{1}^2 + 1318 g_{BY}^3 g_{1}^2+1120 \sqrt{10} g_{BY} g_{YB}^2 g_{1}^2 
                   + 920 g_{BY} g_{YB}^2 g_{1}^2  
       \nonumber\\ 
             &-&200 g_{BY} y_{t}^2 g_{1}^2+ \frac{2240}{3} \sqrt{10} g_{BY} g_{BL}^2 g_{1}^2 
                   +\frac{1840}{3} g_{BY} g_{BL}^2 g_{1}^2+360 \sqrt{10} g_{2}^2 g_{BY} g_{1}^2
        \nonumber\\ 
             &+&  270 g_{2}^2 g_{BY} g_{1}^2+320 \sqrt{10} g_{3}^2 g_{BY} g_{1}^2+880 g_{3}^2 g_{BY} g_{1}^2 
                     + 1120 \sqrt{10} g_{BY}^2 g_{YB} g_{1}^2 
         \nonumber\\     
             &+&3680 g_{BY}^2 g_{YB} g_{1}^2+656 \sqrt{\frac{5}{3}} g_{BY}^2 g_{BL} g_{1}^2
                + 3680 \sqrt{\frac{2}{3}} g_{BY}^2 g_{BL} g_{1}^2 + 328 \sqrt{10} g_{BY}^4 g_{1} 
         \nonumber\\
              &+&920 g_{BY}^4 g_{1}+14400 g_{YB}^4 g_{1} + 360 \sqrt{10} g_{2}^2 g_{BY}^2 g_{1}
                 +1800 g_{2}^2 g_{BY}^2 g_{1} 
          \nonumber\\
              &+& 320 \sqrt{10} g_{3}^2 g_{BY}^2 g_{1}+ 1600 g_{3}^2 g_{BY}^2 g_{1} 
               + 1120 \sqrt{10} g_{BY}^2 g_{YB}^2 g_{1}+ 5600 g_{BY}^2 g_{YB}^2 g_{1}
          \nonumber\\
               &-& 80 g_{BY}^2 y_{t}^2 g_{1} -  85 \sqrt{10} g_{BY} g_{YB} y_{t}^2 g_{1} 
                    +\frac{2240}{3} \sqrt{10} g_{BY}^2 g_{BL}^2 g_{1} 
           \nonumber\\
               &+&\frac{11200}{3} g_{BY}^2 g_{BL}^2 g_{1}+9600 g_{YB}^2 g_{BL}^2 g_{1} 
                   - 45 g_{BY}^2 Y_{N}^2 g_{1}+1120 \sqrt{10} g_{BY}^3 g_{YB} g_{1}
            \nonumber\\ 
               &+& 2240 \sqrt{\frac{5}{3}}g_{BY}^3 g_{BL} g_{1} +3680 \sqrt{\frac{2}{3}} g_{BY}^3 g_{BL} g_{1}
                -100 \sqrt{\frac{5}{3}} g_{BY} y_{t}^2 g_{BL} g_{1} 
    \nonumber \\
        &+& 920 g_{BY}^5+ 14400 g_{BY} g_{YB}^4+1800 g_{2}^2 g_{BY}^3+1600 g_{3}^2 g_{BY}^3+5600 g_{BY}^3 g_{YB}^2 
    \nonumber \\
        &-& 50 \sqrt{10} g_{BY}^2 g_{YB} y_{t}^2+\frac{11200}{3} g_{BY}^3 g_{BL}^2+ 9600 g_{BY} g_{YB}^2 g_{BL}^2   
    \nonumber \\
        &-& 15 \sqrt{15}  g_{BY}^2  g_{BL} Y_{N}^2+ 2240  \sqrt{ \frac{5}{3} } g_{BY}^4 g_{BL} 
             - 80 \sqrt{\frac{5}{3}} g_{BY}^2 y_{t}^2 g_{BL}\Big),   
    \nonumber \\ 
\beta_{g_{2}}^{(2)} &=& \frac{1}{30} g_{2}^3 \Big(4 \sqrt{10} g_{1} g_{YB}+11 g_{1}^2+45 g_{2}^2-260 g_{3}^2
      \nonumber \\
                  &+&  8\sqrt{\frac{5}{3}} g_{BL} g_{BY}+\frac{40 g_{BL}^2}{3}+11 g_{BY}^2+20 g_{YB}^2-20 y_{t}^2\Big), 
      \nonumber\\
\beta_{g_{3}}^{(2)} &=& \frac{1}{10} g_{3}^3 \Big(36 \sqrt{10} g_{1} g_{YB}+27 g_{1}^2+175 g_{2}^2+360 g_{3}^2
      \nonumber\\
                  &+& 24\sqrt{15} g_{BL} g_{BY}+120 g_{BL}^2+27 g_{BY}^2+180 g_{YB}^2-45 y_{t}^2\Big),  
      \nonumber
\eea
\bea
\beta_{g_{BY}}^{(2)} 
        &=&  \frac{1}{100} \Big(18 g_{BY} g_{1}^4+380 g_{YB} g_{1}^4+328 \sqrt{\frac{5}{3}} g_{BL} g_{1}^4
            +18 g_{BY}^2 g_{1}^3+328 \sqrt{10} g_{YB}^2 g_{1}^3
     \nonumber\\
         &+& 164 \sqrt{10} g_{BY} g_{YB} g_{1}^3 + 380 g_{BY} g_{YB} g_{1}^3 
              +328 \sqrt{\frac{5}{3}} g_{BY} g_{BL} g_{1}^3+ 920 \sqrt{\frac{2}{3}} g_{BY} g_{BL} g_{1}^3 
      \nonumber\\ 
          &+&1840 \sqrt{\frac{2}{3}} g_{YB} g_{BL}g_{1}^3+ 18 g_{BY}^3 g_{1}^2+920 g_{YB}^3 g_{1}^2
                +\frac{2240}{3} \sqrt{\frac{5}{3}} g_{BL}^3 g_{1}^2+ 328 \sqrt{10} g_{BY} g_{YB}^2 g_{1}^2
      \nonumber\\
           &+& 1840 g_{BY} g_{YB}^2 g_{1}^2+ \frac{3680}{3} g_{BY} g_{BL}^2 g_{1}^2
                +\frac{1840}{3} g_{YB} g_{BL}^2 g_{1}^2 + 180 g_{2}^2 g_{YB} g_{1}^2+880 g_{3}^2 g_{YB}g_{1}^2
      \nonumber \\
           &+& 380 g_{BY}^2 g_{YB} g_{1}^2-85\sqrt{10} y_{t}^{\ast} g_{YB} y_{t} g_{1}^2
                + 320\sqrt{\frac{5}{3}} g_{3}^2 g_{BL} g_{1}^2+328 \sqrt{\frac{5}{3}} g_{BY}^2 g_{BL} g_{1}^2
      \nonumber \\
          &+& 920 \sqrt{\frac{2}{3}} g_{BY}^2 g_{BL} g_{1}^2+1120 \sqrt{\frac{5}{3}} g_{YB}^2 g_{BL}g_{1}^2 
               + 2896 \sqrt{\frac{5}{3}} g_{BY} g_{YB} g_{BL} g_{1}^2+1840 \sqrt{\frac{2}{3}} g_{BY}g_{YB} g_{BL} g_{1}^2
      \nonumber \\
          &-&100 \sqrt{\frac{5}{3}} y_{t}^{\ast} y_{t} g_{BL} g_{1}^2+18g_{BY}^4 g_{1}+ 920 g_{BY} g_{YB}^3 g_{1}
               +\frac{2240}{3} \sqrt{\frac{5}{3}} g_{BY} g_{BL}^3 g_{1} 
     \nonumber \\
        &+& \frac{11200}{3} \sqrt{\frac{2}{3}} g_{BY}g_{BL}^3 g_{1}+3200 \sqrt{6} g_{YB} g_{BL}^3 g_{1}
         + 90 g_{2}^2 g_{BY}^2g_{1}+1840 g_{BY}^2 g_{YB}^2 g_{1}
      \nonumber \\
        &+& \frac{2240}{3} \sqrt{10} g_{BY}^2 g_{BL}^2g_{1}+\frac{1120}{3} \sqrt{10} g_{BY} g_{YB}g_{BL}^2 g_{1}
               + \frac{1840}{3} g_{BY} g_{YB} g_{BL}^2 g_{1}-60 \sqrt{15}g_{BY} g_{BL} Y_{N}^2 g_{1}
     \nonumber \\
        &+& 164 \sqrt{10} g_{BY}^3 g_{YB} g_{1}+180 \sqrt{10} g_{2}^2 g_{BY} g_{YB} g_{1}
           +180 g_{2}^2 g_{BY} g_{YB} g_{1}+880 g_{3}^2 g_{BY} g_{YB} g_{1}
     \nonumber \\
       &-& 50 \sqrt{10} y_{t}^{\ast} g_{BY} g_{YB} y_{t} g_{1}+328\sqrt{\frac{5}{3}} g_{BY}^3 g_{BL} g_{1}
            +920 \sqrt{\frac{2}{3}} g_{BY}^3 g_{BL} g_{1}
     \nonumber \\
       &+& 1120 \sqrt{\frac{5}{3}} g_{BY} g_{YB}^2 g_{BL}g_{1}+5600 \sqrt{\frac{2}{3}} g_{BY} g_{YB}^2 g_{BL} g_{1}
          + 120 \sqrt{15} g_{2}^2 g_{BY} g_{BL} g_{1}
     \nonumber \\
        &+& 600 \sqrt{6} g_{2}^2 g_{BY} g_{BL} g_{1}+320\sqrt{\frac{5}{3}} g_{3}^2 g_{BY} g_{BL} g_{1}
          +1600 \sqrt{\frac{2}{3}} g_{3}^2 g_{BY}g_{BL} g_{1}
     \nonumber \\
       &+& 2896 \sqrt{\frac{5}{3}} g_{BY}^2 g_{YB} g_{BL} g_{1}+1840\sqrt{\frac{2}{3}} g_{BY}^2 g_{YB} g_{BL} g_{1}
            -  80 \sqrt{\frac{5}{3}} y_{t}^{\ast} g_{BY} y_{t} g_{BL} g_{1}+560 \sqrt{10} g_{BY}^2 g_{YB}^3
     \nonumber \\
       &+& \frac{11200}{3} \sqrt{\frac{2}{3}} g_{BY}^2 g_{BL}^3+3200 \sqrt{6} g_{BY} g_{YB} g_{BL}^3
           +1120 \sqrt{10} g_{BY}^3 g_{BL}^2-\frac{400}{3} y_{t}^{\ast} g_{BY} y_{t} g_{BL}^2
     \nonumber \\
       &-& 300 g_{BY} g_{BL}^2 Y_{N}^2+180 \sqrt{10} g_{2}^2 g_{BY}^2 g_{YB} 
          + 160 \sqrt{10} g_{3}^2 g_{BY}^2 g_{YB}-425 y_{t}^{\ast} g_{BY} g_{YB}^2 y_{t} 
     \nonumber \\
       &+& 4800 \sqrt{6} g_{BY} g_{YB}^3 g_{BL}+600 \sqrt{6} g_{2}^2 g_{BY}^2 g_{BL}
         + 1600 \sqrt{\frac{2}{3}} g_{3}^2 g_{BY}^2 g_{BL}+5600 \sqrt{\frac{2}{3}} g_{BY}^2g_{YB}^2 g_{BL}
   \nonumber \\
       &+& 1840 \sqrt{\frac{2}{3}} g_{BY}^3 g_{YB} g_{BL}-500 \sqrt{\frac{2}{3}} y_{t}^{\ast} g_{BY} g_{YB} y_{t} g_{BL} 
          +120 \sqrt{15} g_{2}^2 g_{BL} g_{1}^2+560 \sqrt{10} g_{BY} g_{YB}^3 g_{1}
   \nonumber \\
       &+& 160 \sqrt{10} g_{3}^2 g_{BY} g_{YB} g_{1}+4800 \sqrt{6} g_{YB}^3 g_{BL} g_{1} 
          +380 g_{BY}^3 g_{YB} g_{1}+\frac{3680}{3} g_{BY}^2 g_{BL}^2 g_{1}
   \nonumber \\
       &+& 164 \sqrt{10} g_{BY}^4 g_{YB}+90 g_{2}^2 g_{BY} g_{1}^2 
             +164 \sqrt{10} g_{BY}^2 g_{YB} g_{1}^2+920 \sqrt{\frac{2}{3}} g_{BY}^4 g_{BL} \Big),  
   \nonumber
\eea
\bea
\beta_{g_{YB}}^{(2)} 
  &=& \frac{1}{100} \Big(920 g_{1} g_{YB}^4+560 \sqrt{10} g_{BY} g_{YB}^4+4800 \sqrt{6} g_{BL} g_{YB}^4
      + 328 \sqrt{10} g_{1}^2 g_{YB}^3+9600 g_{BL}^2 g_{YB}^3
  \nonumber \\
    &+& 1840 g_{1} g_{BY} g_{YB}^3 + 1120 \sqrt{\frac{5}{3}} g_{1} g_{BL} g_{YB}^3
      +920 \sqrt{\frac{2}{3}} g_{1}g_{BL} g_{YB}^3+ 1120 \sqrt{\frac{5}{3}} g_{BY} g_{BL} g_{YB}^3
  \nonumber \\
    &+&5600 \sqrt{\frac{2}{3}} g_{BY} g_{BL} g_{YB}^3+ 380 g_{1}^3g_{YB}^2+164 \sqrt{10} g_{BY}^3 g_{YB}^2
      +3200 \sqrt{6} g_{BL}^3 g_{YB}^2
   \nonumber \\
       &+& 180 g_{1} g_{2}^2 g_{YB}^2+880 g_{1} g_{3}^2 g_{YB}^2+380 g_{1} g_{BY}^2 g_{YB}^2
         + \frac{1120}{3} \sqrt{10} g_{1} g_{BL}^2 g_{YB}^2+\frac{1840}{3} g_{1} g_{BL}^2g_{YB}^2
    \nonumber \\
       &+& \frac{1120}{3} \sqrt{10} g_{BY} g_{BL}^2 g_{YB}^2+\frac{11200}{3} g_{BY} g_{BL}^2 g_{YB}^2
          + 164 \sqrt{10} g_{1}^2 g_{BY}g_{YB}^2+180 \sqrt{10} g_{2}^2 g_{BY} g_{YB}^2
   \nonumber \\
       &+& 160 \sqrt{10} g_{3}^2 g_{BY} g_{YB}^2+656 \sqrt{\frac{5}{3}} g_{1}^2 g_{BL}g_{YB}^2
           +1840 \sqrt{\frac{2}{3}} g_{1}^2 g_{BL} g_{YB}^2+1840 \sqrt{\frac{2}{3}} g_{BY}^2 g_{BL} g_{YB}^2
    \nonumber \\
       &+& 2896 \sqrt{\frac{5}{3}} g_{1}g_{BY} g_{BL} g_{YB}^2+1840 \sqrt{\frac{2}{3}} g_{1} g_{BY} g_{BL} g_{YB}^2
           + 6400 g_{BL}^4 g_{YB}+18 g_{1}g_{BY}^3 g_{YB}
    \nonumber \\
       &+&\frac{2240}{3} \sqrt{\frac{5}{3}} g_{1} g_{BL}^3 g_{YB}+ \frac{1840}{3} \sqrt{\frac{2}{3}} g_{1} g_{BL}^3 g_{YB}
          +\frac{2240}{3} \sqrt{\frac{5}{3}} g_{BY} g_{BL}^3 g_{YB}
    \nonumber \\
       &+& \frac{11200}{3} \sqrt{\frac{2}{3}} g_{BY} g_{BL}^3 g_{YB}+\frac{3680}{3} g_{1}^2 g_{BL}^2 g_{YB}
         + \frac{2240}{3} \sqrt{10} g_{BY}^2 g_{BL}^2 g_{YB}+\frac{3680}{3} g_{BY}^2 g_{BL}^2 g_{YB}
    \nonumber \\
       &+& \frac{2896}{3} \sqrt{10} g_{1} g_{BY} g_{BL}^2 g_{YB}+\frac{3680}{3} g_{1} g_{BY} g_{BL}^2 g_{YB}
           + 18 g_{1}^3 g_{BY} g_{YB}+90 g_{1} g_{2}^2 g_{BY} g_{YB}
    \nonumber \\
       &-& 80 y_{t}^{\ast} g_{BY}^2 y_{t} g_{YB}-200 y_{t}^{\ast} g_{1} g_{BY} y_{t} g_{YB}
           + 328 \sqrt{\frac{5}{3}} g_{1}^3 g_{BL} g_{YB}+380 \sqrt{\frac{2}{3}} g_{1}^3 g_{BL}g_{YB}
    \nonumber \\
       &+& 328 \sqrt{\frac{5}{3}} g_{BY}^3 g_{BL} g_{YB}+920 \sqrt{\frac{2}{3}} g_{BY}^3 g_{BL} g_{YB}
           +120 \sqrt{15} g_{1} g_{2}^2 g_{BL} g_{YB}+60 \sqrt{6} g_{1} g_{2}^2 g_{BL} g_{YB}
    \nonumber \\
       &+& 320 \sqrt{\frac{5}{3}} g_{1} g_{3}^2 g_{BL} g_{YB}+880\sqrt{\frac{2}{3}} g_{1} g_{3}^2 g_{BL} g_{YB}
           + 328 \sqrt{\frac{5}{3}} g_{1} g_{BY}^2 g_{BL} g_{YB}+380 \sqrt{\frac{2}{3}} g_{1} g_{BY}^2 g_{BL} g_{YB}
     \nonumber \\
       &+& 328 \sqrt{\frac{5}{3}} g_{1}^2 g_{BY} g_{BL} g_{YB}+920 \sqrt{\frac{2}{3}} g_{1}^2 g_{BY} g_{BL}g_{YB}
           +120 \sqrt{15} g_{2}^2 g_{BY} g_{BL} g_{YB}
    \nonumber \\
       &+&600 \sqrt{6} g_{2}^2 g_{BY} g_{BL} g_{YB} 
 + 320 \sqrt{\frac{5}{3}} g_{3}^2 g_{BY} g_{BL} g_{YB}+1600 \sqrt{\frac{2}{3}} g_{3}^2 g_{BY} g_{BL} g_{YB} 
    \nonumber \\
        &-& 170 \sqrt{\frac{5}{3}} y_{t}^{\ast} g_{1} y_{t} g_{BL} g_{YB}
            - 100 \sqrt{\frac{5}{3}} y_{t}^{\ast} g_{BY} y_{t} g_{BL} g_{YB}-180 Y_{N} g_{BY}^2 Y_{N} g_{YB}
 \nonumber \\
       &+& \frac{2240}{9}\sqrt{10} g_{1} g_{BL}^4+\frac{22400}{9} g_{BY} g_{BL}^4
            +\frac{4480}{3} \sqrt{\frac{5}{3}} g_{BY}^2 g_{BL}^3+ \nonumber \\
       &+& \frac{3680}{3} \sqrt{\frac{2}{3}} g_{1} g_{BY} g_{BL}^3+\frac{328}{3} \sqrt{10} g_{1}^3 g_{BL}^2
          + \frac{1840}{3} g_{BY}^3 g_{BL}^2+120 \sqrt{10} g_{1} g_{2}^2 g_{BL}^2
   \nonumber \\
       &+& \frac{320}{3} \sqrt{10} g_{1} g_{3}^2 g_{BL}^2+\frac{328}{3} \sqrt{10} g_{1} g_{BY}^2 g_{BL}^2 
          +  \frac{1840}{3} g_{1}^2 g_{BY} g_{BL}^2+1200 g_{2}^2 g_{BY} g_{BL}^2
   \nonumber \\
       &+& \frac{3200}{3} g_{3}^2 g_{BY} g_{BL}^2-\frac{100}{3} \sqrt{10} y_{t}^{\ast} g_{1} y_{t} g_{BL}^2
          -  \frac{80}{3} \sqrt{10} y_{t}^{\ast} g_{BY} y_{t} g_{BL}^2+6 \sqrt{6} g_{1} g_{BY}^3 g_{BL}
   \nonumber \\
       &+& 6\sqrt{6} g_{1}^3 g_{BY} g_{BL}+30 \sqrt{6} g_{1} g_{2}^2 g_{BY} g_{BL}
          - 60\sqrt{10} Y_{N} g_{BY} g_{BL}^2 Y_{N}-170 y_{t}^{\ast} g_{1}^2 y_{t} g_{YB}\Big), 
   \nonumber
\eea
\bea
\beta_{g_{BL}}^{(2)} 
     &=& \sqrt{\frac{3}{2}}\frac{1}{100} \Big(920 g_{YB}^5+328 \sqrt{10} g_{1} g_{YB}^4+2240 \sqrt{\frac{5}{3}} g_{BL} g_{YB}^4
        + 920\sqrt{\frac{2}{3}} g_{BL} g_{YB}^4+380 g_{1}^2 g_{YB}^3
 \nonumber \\
      &+& 380 g_{BY}^2 g_{YB}^3+\frac{2240}{3} \sqrt{10} g_{BL}^2 g_{YB}^3+\frac{41840}{3} g_{BL}^2 g_{YB}^3
         + 656\sqrt{\frac{5}{3}} g_{1} g_{BL} g_{YB}^3
  \nonumber \\
      &+&3680 \sqrt{\frac{2}{3}} g_{1} g_{BL} g_{YB}^3+ 656\sqrt{\frac{5}{3}} g_{BY} g_{BL} g_{YB}^3
           +\frac{4480}{3} \sqrt{\frac{5}{3}} g_{BL}^3 g_{YB}^2 + \frac{41840}{3} \sqrt{\frac{2}{3}} g_{BL}^3 g_{YB}^2
 \nonumber \\  
       &+&\frac{2240}{3} \sqrt{10} g_{1} g_{BL}^2 g_{YB}^2+\frac{7360}{3} g_{1} g_{BL}^2 g_{YB}^2
         +\frac{656}{3} \sqrt{10} g_{BY} g_{BL}^2 g_{YB}^2+ \frac{7360}{3} g_{BY} g_{BL}^2 g_{YB}^2
 \nonumber \\
        &-&85 \sqrt{10} y_{t}^{\ast} g_{1} y_{t} g_{YB}^2- 50 \sqrt{10} y_{t}^{\ast} g_{BY}y_{t} g_{YB}^2
          +656 \sqrt{\frac{5}{3}} g_{1}^2 g_{BL} g_{YB}^2+ 380 \sqrt{\frac{2}{3}} g_{1}^2 g_{BL} g_{YB}^2
 \nonumber \\
        &+& 240 \sqrt{15} g_{2}^2 g_{BL} g_{YB}^2+ 60 \sqrt{6} g_{2}^2 g_{BL} g_{YB}^2
          +640 \sqrt{\frac{5}{3}} g_{3}^2 g_{BL} g_{YB}^2+  880 \sqrt{\frac{2}{3}} g_{3}^2 g_{BL} g_{YB}^2
 \nonumber \\
        &+&656 \sqrt{\frac{5}{3}} g_{BY}^2 g_{BL} g_{YB}^2+ 380 \sqrt{\frac{2}{3}} g_{BY}^2 g_{BL} g_{YB}^2
          -425 \sqrt{\frac{2}{3}} y_{t}^{\ast} y_{t} g_{BL} g_{YB}^2+ 18 g_{BY}^4 g_{YB}
 \nonumber \\
       &+&\frac{4480}{9} \sqrt{10} g_{BL}^4 g_{YB}+\frac{80000}{9} g_{BL}^4 g_{YB}
         + \frac{4480}{3} \sqrt{\frac{5}{3}} g_{1} g_{BL}^3 g_{YB}+\frac{4480}{3} \sqrt{\frac{5}{3}} g_{BY} g_{BL}^3 g_{YB}
 \nonumber \\
      &+& \frac{7360}{3} \sqrt{\frac{2}{3}} g_{BY} g_{BL}^3 g_{YB}+18 g_{1}^2 g_{BY}^2 g_{YB}
         +90 g_{2}^2 g_{BY}^2 g_{YB}+\frac{656}{3} \sqrt{10} g_{1}^2 g_{BL}^2 g_{YB}
 \nonumber \\         
      &+& \frac{1840}{3} g_{1}^2 g_{BL}^2 g_{YB} + 240 \sqrt{10} g_{2}^2 g_{BL}^2 g_{YB}+1200 g_{2}^2 g_{BL}^2 g_{YB}
         + \frac{640}{3} \sqrt{10} g_{3}^2 g_{BL}^2 g_{YB}
 \nonumber \\      
      &+&\frac{3200}{3} g_{3}^2 g_{BL}^2 g_{YB}+\frac{656}{3} \sqrt{10} g_{BY}^2 g_{BL}^2 g_{YB}
          +\frac{1840}{3} g_{BY}^2 g_{BL}^2 g_{YB}-\frac{1000}{3} y_{t}^{\ast} y_{t}  g_{BL}^2 g_{YB}
 \nonumber \\
      &-&100 \sqrt{\frac{5}{3}} y_{t}^{\ast} g_{1} y_{t} g_{BL} g_{YB}- 80 \sqrt{\frac{5}{3}} y_{t}^{\ast} g_{BY} y_{t} g_{BL} g_{YB}
        -60 \sqrt{15} Y_{N} g_{BY} g_{BL} Y_{N} g_{YB}
 \nonumber \\
      &+&\frac{80000}{9} \sqrt{\frac{2}{3}} g_{BL}^5+\frac{4480}{9} \sqrt{10} g_{BY} g_{BL}^4
        + \frac{1840}{3}\sqrt{\frac{2}{3}} g_{1}^2 g_{BL}^3+400 \sqrt{6} g_{2}^2 g_{BL}^3
 \nonumber \\
    &+&\frac{3200}{3} \sqrt{\frac{2}{3}} g_{3}^2 g_{BL}^3+\frac{1840}{3} \sqrt{\frac{2}{3}} g_{BY}^2 g_{BL}^3
       -\frac{400}{3} \sqrt{\frac{2}{3}} y_{t}^{\ast} y_{t} g_{BL}^3+6 \sqrt{6} g_{BY}^4 g_{BL}
       +6 \sqrt{6} g_{1}^2 g_{BY}^2 g_{BL}
  \nonumber \\
      &+& 30 \sqrt{6} g_{2}^2 g_{BY}^2 g_{BL}-100 \sqrt{6} Y_{N} g_{BL}^3 Y_{N} 
         +180 g_{2}^2 g_{YB}^3+880 g_{3}^2 g_{YB}^3\Big) . 
\eea
\subsection{The two-loop beta function for the top Yukawa coupling}
\bea
\beta_{y_{t}}^{(2)} 
   &=& \frac{1}{600} y_{t} \Big(3600 \Big(\frac{\lambda}{2}\Big)^2+500 \sqrt{15} g_{BY} y_{t} g_{BL}y_{t}^{\ast}
        + 1000 y_{t} g_{BL}^2 y_{t}^{\ast}-340\sqrt{\frac{2}{3}} g_{1} g_{BY} g_{YB} g_{BL}
   \nonumber \\
        &+& 50 \sqrt{\frac{5}{3}} g_{1}^2 g_{BY} g_{BL}-\frac{100}{3} \sqrt{10} g_{1} g_{YB} g_{BL}^2
             + \frac{190}{3} g_{1}^2 g_{BL}^2+270 \sqrt{15} g_{2}^2 g_{BY} g_{BL}
       \nonumber \\
        &+&450 g_{2}^2  \sqrt{\frac{5}{3}} g_{BL}^2-800g_{3}^2 g_{BY} g_{BL}
          -\frac{1600}{3} g_{3}^2 g_{BL}^2-100 \sqrt{\frac{5}{3}} g_{BY}  g_{YB}^2 g_{BL}
     \nonumber \\
        &+&\frac{26600}{3} \sqrt{\frac{5}{3}} g_{BY} g_{BL}^3+10850 g_{BY}^2 g_{BL}^2
           +4016 \sqrt{\frac{5}{3}} g_{BY}^3 g_{BL}- \frac{200}{3} g_{YB}^2 g_{BL}^2
           +\frac{40600}{9} g_{BL}^4
    \nonumber \\
       &+&750 \sqrt{10} g_{1} g_{YB} y_{t} y_{t}^{\ast}+1275 g_{1}^2 y_{t} y_{t}^{\ast}
            +3375 g_{2}^2 y_{t} y_{t}^{\ast}+12000 g_{3}^2  y_{t} y_{t}^{\ast}+1275 g_{BY}^2 y_{t} y_{t}^{\ast}
     \nonumber \\
          &+&1500 g_{YB}^2 y_{t} y_{t}^{\ast}- 4050 y_{t}^2 \Big(y_{t}^{\ast}\Big)^2+405\sqrt{10} g_{1} g_{2}^2 g_{YB}
              -270 g_{1}^2 g_{2}^2- 400 \sqrt{10} g_{1} g_{3}^2 g_{YB}
    \nonumber \\
          &+&760 g_{1}^2 g_{3}^2+25 \sqrt{10}g_{1} g_{BY}^2 g_{YB}- 106 g_{1}^2 g_{BY}^2
            +2008 \sqrt{10}g_{1}^3 g_{YB}+ 16275 g_{1}^2 g_{YB}^2
   \nonumber \\
      &+& 6650 \sqrt{10} g_{1} g_{YB}^3+1187 g_{1}^4+5400 g_{2}^2g_{3}^2-270 g_{2}^2 g_{BY}^2
            +675 g_{2}^2 g_{YB}^2-3450 g_{2}^4+760 g_{3}^2 g_{BY}^2
    \nonumber \\
       &-&800 g_{3}^2 g_{YB}^2- 64800 g_{3}^4+ 95 g_{BY}^2 g_{YB}^2+1187 g_{BY}^4
          +10150 g_{YB}^4+300 \lambda_{3}^2\Big)
    \nonumber \\
      &+& \frac{320}{3} g_{BL}^2+\frac{1}{80} \Big(y_{t}^2 y_{t}^{\ast} \Big(-960 \frac{\lambda }{2}
          +100 \sqrt{\frac{5}{3}} g_{BY} g_{BL}+-540 y_{t} y_{t}^{\ast}+50 \sqrt{10} g_{1}g_{YB} 
    \nonumber \\
      &+& 223 g_{1}^2+675 g_{2}^2+1280 g_{3}^2+223 g_{BY}^2+160 g_{YB}^2\Big) +120 y_{t}^3 \Big(y_{t}^{\ast}\Big)^2\Big) .
\eea

\subsection{The two-loop beta functions for the heavy neutrino Yukawa coupling}
\bea
\beta_{Y_{N}}^{(2)} 
    &=& \frac{1}{20} \Big(-Y_{N} \Big(-225 Y_{N} \Big(\frac{2}{3} g_{BL}^2+g_{YB}^2\Big) Y_{N}
         +128 \sqrt{\frac{5}{3}} g_{BY} g_{BL}^3+70 g_{BY}^2 g_{BL}^2
  \nonumber \\
      &+&7380 g_{YB}^2 g_{BL}^2+ 2540 g_{BL}^4+45 Y_{N}^4 +96 \sqrt{10} g_{1} g_{YB}^3
       +105 g_{1}^2 g_{YB}^2+5715 g_{YB}^4 
   \nonumber \\
      &-& 80 \lambda_{2}^2-20 \lambda_{3}^2\Big)+880 g_{BL}^2 Y_{N}^2 Y_{N}
       + 1320 g_{YB}^2 Y_{N}^2 Y_{N}-160 \lambda_{2} Y_{N}^2 Y_{N}- 10 Y_{N}^5  .
\eea
\subsection{The two-loop beta functions for the scalar quartic couplings}
\bea
\beta_{\lambda}^{(2)} 
    &=& 2\Big(-\frac{3411 g_{1}^6}{2000}-\frac{4221 g_{BY}^2 g_{1}^4}{2000}+\frac{1887}{400} \lambda g_{1}^4
         -\frac{48}{5} \sqrt{\frac{2}{5}} g_{2}^2 g_{YB} g_{1}^3+12 \sqrt{\frac{2}{5}} g_{YB} \lambda g_{1}^3
   \nonumber \\
     &-& \frac{10971 g_{BY}^4 g_{1}^2}{2000}-\frac{8}{5} \Big(y_{t}^{\ast}\Big)^2 y_{t}^2 g_{1}^2
        + \frac{63}{10}y_{t}^{\ast} g_{2}^2 y_{t} g_{1}^2+\frac{621}{200} g_{BY}^2 \lambda g_{1}^2
  \nonumber \\
     &+&  \frac{153}{10} g_{YB}^2 \lambda g_{1}^2+18 g_{YB}^2 \lambda_{3} g_{1}^2
            -  2 \sqrt{10} \Big(y_{t}^{\ast}\Big)^2 g_{YB} y_{t}^2 g_{1}+5 \sqrt{\frac{5}{2}} y_{t}^{\ast} g_{YB} y_{t} \lambda g_{1}
  \nonumber \\
      &-&\frac{12}{5} \sqrt{6} y_{t}^{\ast} g_{BY} g_{YB} y_{t} g_{BL} g_{1}+12 \sqrt{6} g_{BY} g_{YB} \lambda_{3} g_{BL} g_{1}
            + \frac{305 g_{2}^6}{16}- \frac{4221 g_{BY}^6}{2000} 
   \nonumber \\          
      &-&39 \lambda^3-4 \lambda_{3}^3-32 \Big(y_{t}^{\ast}\Big)^2 g_{3}^2 y_{t}^2 
          -\frac{8}{5} \Big(y_{t}^{\ast}\Big)^2 g_{BY}^2 y_{t}^2-4 \Big(y_{t}^{\ast}\Big)^2 g_{YB} y_{t}^2
   \nonumber \\
     &+& 27 g_{2}^2 \lambda^2-36 y_{t}^{\ast} y_{t} \lambda^2+g_{YB}^2 \lambda_{3}^2
        - 5 \lambda\lambda_{3}^2-\frac{8}{3} \Big(y_{t}^{\ast}\Big)^2 y_{t}^2 g_{BL}^2+32 \lambda_{3}^2 g_{BL}^2
   \nonumber \\      
        &+& \frac{51}{5} g_{BY}^2 \lambda g_{BL}^2+12 g_{BY}^2 \lambda_{3} g_{BL}^2
          - \frac{289}{80} \Big(g_{1}^2 g_{2}^4+g_{BY}^2 g_{2}^4\Big)-\frac{1677}{400} \Big(g_{2}^2 g_{1}^4
            + g_{2}^2 g_{BY}^4\Big)
    \nonumber \\         
         &-& \frac{9}{4} y_{t}^{\ast} g_{2}^4 y_{t}+\frac{63}{10}y_{t}^{\ast} g_{2}^2 g_{BY}^2 y_{t}
           -\frac{171}{100} y_{t}^{\ast} \Big(g_{1}^4+2 g_{BY}^2 g_{1}^2+ g_{BY}^4\Big) y_{t}
           -\frac{73}{16}g_{2}^4 \lambda+\frac{1887}{400} g_{BY}^4 \lambda
   \nonumber \\
       &-& \frac{3}{2} \Big(y_{t}^{\ast}\Big)^2 y_{t}^2 \lambda+40 y_{t}^{\ast} g_{3}^3 y_{t} \lambda 
             +\frac{45}{4} y_{t}^{\ast} g_{2}^2 y_{t} \lambda+5 y_{t}^{\ast} g_{YB}^2 y_{t} \lambda
               + \frac{108}{5} \Big(\frac{1}{4} g_{1}^2 \lambda ^2+\frac{1}{4} g_{BY}^2 \lambda^2\Big)
   \nonumber \\             
      &+& \frac{117}{20} \Big(\frac{1}{2} g_{1}^2 \lambda g_{2}^2+g_{BY}^2 \lambda_{3} g_{2}^2\Big)
        -\frac{32}{5} \sqrt{\frac{3}{5}} g_{2}^2 g_{BY}^3 g_{BL}-4 \sqrt{\frac{5}{3}} \Big(y_{t}^{\ast}\Big)^2 g_{BY} y_{t}^2 g_{BL}
 \nonumber \\
      &+& 8 \sqrt{\frac{3}{5}} g_{BY}^3 \lambda g_{BL}+5 \sqrt{\frac{2}{3}} y_{t}^{\ast} y_{t} \lambda g_{BL}
            + 5 \sqrt{\frac{5}{3}} y_{t}^{\ast} g_{BY} y_{t} \lambda g_{BL} 
 \nonumber \\
       &-& 9 \sqrt{\frac{2}{5}} y_{t}^{\ast} y_{t} \Big(g_{YB} g_{1}^3
             +\sqrt{\frac{2}{3}} g_{BY} g_{BL}  g_{1}^2-g_{2}^2 g_{YB} g_{1}+g_{BY}^2 g_{YB} g_{1}
        -\sqrt{\frac{2}{3}} g_{2}^2 g_{BY} g_{BL}\Big)
 \nonumber \\     
     &-&\frac{144}{25} \sqrt{\frac{2}{5}} \Big(g_{YB} g_{1}^5+g_{BY}^2  g_{YB} g_{1}^3
        +\sqrt{\frac{2}{3}} g_{BY}^3 g_{BL} g_{1}^2+\sqrt{\frac{2}{3}} g_{BY}^5 g_{BL}\Big)
 \nonumber \\
     &+& \frac{17}{4} y_{t}^{\ast} y_{t}\lambda \Big(g_{1}^2+\frac{2}{3} g_{BL}^2\Big)
         -\frac{18}{5} y_{t}^{\ast} y_{t} \Big(g_{1}^2 g_{YB}^2+\frac{2}{3} g_{BY}^2 g_{BL}^2\Big) 
         -\frac{117}{10} \Big(g_{1}^2 g_{2}^2 g_{YB}^2+\frac{2}{3} g_{2}^2 g_{BY}^2 g_{BL}^2\Big) 
 \nonumber \\
      &-&  \frac{351}{50} \Big(g_{YB}^2 g_{1}^4+g_{BY}^2 g_{YB}^2 g_{1}^2+\frac{2}{3} g_{BY}^2 g_{BL}^2 g_{1}^2
        +\frac{2}{3} g_{BY}^4 g_{BL}^2\Big)\Big), 
 \nonumber
\eea
\bea
\beta_{\lambda_{2}}^{(2)} 
   &=& \frac{2}{5} \Big(1440 g_{YB}^2 g_{BL}^2 Y_{N} Y_{N}+75 \lambda_{2} g_{BL}^2 Y_{N} Y_{N} 
        +90 g_{BL}^2 Y_{N}^2 \Big(Y_{N}\Big)^2+480 g_{BL}^4 Y_{N} Y_{N} 
  \nonumber \\
      &+& 60 \sqrt{6} g_{1} g_{BY} g_{YB} \lambda_{3} g_{BL}-6144 \sqrt{10} g_{1} g_{YB}^3 g_{BL}^2
        -4008 g_{1}^2 g_{YB}^2 g_{BL}^2-4008 g_{BY}^2 g_{YB}^2 g_{BL}^2 
 \nonumber \\
     &+& 640 \sqrt{\frac{5}{3}} g_{BY} \lambda_{2} g_{BL}^3+422 g_{BY}^2 \lambda_{2} g_{BL}^2
         + 60 g_{BY}^2 \lambda_{3} g_{BL}^2-4096 \sqrt{\frac{5}{3}} g_{BY} g_{BL}^5-2672 g_{BY}^2 g_{BL}^4
  \nonumber \\        
     &+& 11760 g_{YB}^2 \lambda_{2} g_{BL}^2-83520 g_{YB}^4 g_{BL}^2-55680 g_{YB}^2  g_{BL}^4
         +5280 \lambda_{2} g_{BL}^4+1120 \lambda_{2}^2 g_{BL}^2-17920 g_{BL}^6
  \nonumber \\
     &+& \frac{225}{2} g_{YB}^2 \lambda_{2} Y_{N}^2 +1080 g_{YB}^4 Y_{N}^2 
         + 135 g_{YB}^2 Y_{N}^2 \Big(Y_{N}\Big)^2+\frac{45}{2} \lambda_{2} Y_{N}^4 
     -150 \lambda_{2}^2 Y_{N}^2 +270 Y_{N}^6 
   \nonumber \\
     &-& 30 \lambda_{3}^2 y_{t} y_{t}^{\ast}+480 \sqrt{10} g_{1} g_{YB}^3 \lambda_{2}
      +633 g_{1}^2 g_{YB}^2 \lambda_{2}+90 g_{1}^2 g_{YB}^2 \lambda_{3}-4608 \sqrt{10} g_{1} g_{YB}^5
   \nonumber \\
      &-&  6012 g_{1}^2 g_{YB}^4+6 g_{1}^2 \lambda_{3}^2+30 g_{2}^2 \lambda_{3}^2
      + 6 g_{BY}^2 \lambda_{3}^2+11880 g_{YB}^4 \lambda_{2}+1680 g_{YB}^2 \lambda_{2}^2
    \nonumber \\
     &-& 60480 g_{YB}^6-50 \lambda_{2} \lambda_{3}^2-600 \lambda_{2}^3-20 \lambda_{3}^3\Big),  
 \nonumber 
\eea
\bea
\beta_{\lambda_{3}}^{(2)} 
   &=& -\frac{6417}{50} g_{YB}^2 g_{1}^4+\frac{1671}{400} \lambda_{3} g_{1}^4
       -\frac{2304}{5} \sqrt{\frac{2}{5}} g_{YB}^3 g_{1}^3+12 \sqrt{\frac{2}{5}} g_{YB} \lambda_{3} g_{1}^3
 \nonumber \\
    &-& \frac{1137}{25} \sqrt{6} g_{BY} g_{YB} g_{BL} g_{1}^3-\frac{4428}{5} g_{YB}^4 g_{1}^2 
          -\frac{81}{2} g_{2}^2 g_{YB}^2 g_{1}^2-\frac{81}{10} g_{BY}^2 g_{YB}^2 g_{1}^2 
 \nonumber \\
    &+& \frac{3}{5} \lambda_{3}^2 g_{1}^2-\frac{27}{5} g_{BY}^2 g_{BL}^2 g_{1}^2
          -216 g_{YB}^2 g_{BL}^2 g_{1}^2-\frac{342}{5} y_{t}^{\ast} g_{YB}^2 y_{t} g_{1}^2
           +108 g_{YB}^2 \lambda  g_{1}^2
   \nonumber\\ 
     &+&72 g_{YB}^2 \lambda_{2} g_{1}^2+\frac{9}{8} g_{2}^2 \lambda_{3} g_{1}^2
       +\frac{81}{40} g_{BY}^2 \lambda_{3} g_{1}^2+\frac{1491}{10} g_{YB}^2 \lambda_{3} g_{1}^2
         +\frac{36}{5} \lambda  \lambda_{3}  g_{1}^2
   \nonumber\\       
       &-& \frac{44608 g_{BY} g_{YB}^2 g_{BL} g_{1}^2}{5\sqrt{15}}-\frac{1424}{5} \sqrt{6} g_{BY} g_{YB} g_{BL}^3 g_{1}
          -\frac{1536}{5} \sqrt{\frac{2}{5}} g_{BY}^2 g_{YB} g_{BL}^2 g_{1}
   \nonumber\\ 
       &-& 48 \sqrt{10} y_{t}^{\ast} g_{YB} y_{t} g_{BL}^2 g_{1}-72 \sqrt{10} y_{t}^{\ast} g_{YB}^3 y_{t} g_{1}
           + 96 \sqrt{10} g_{YB}^3 \lambda_{3} g_{1}-\frac{2016}{5} \sqrt{6} g_{BY} g_{YB}^3 g_{BL} g_{1}
   \nonumber \\
    &-& \frac{1137}{25} \sqrt{6} g_{BY}^3 g_{YB} g_{BL} g_{1} 
          -27 \sqrt{6} g_{2}^2 g_{BY} g_{YB}  g_{BL} g_{1}-\frac{228}{5} \sqrt{6} y_{t}^{\ast} g_{BY} g_{YB} y_{t} g_{BL} g_{1}
    \nonumber\\
     &+&72 \sqrt{6} g_{BY} g_{YB} \frac{\lambda }{2} g_{BL} g_{1}+48 \sqrt{6} g_{BY} g_{YB} \lambda_{2} g_{BL} g_{1}
         +\frac{24}{5} \sqrt{6} g_{BY} g_{YB} \lambda_{3} g_{BL} g_{1} 
   \nonumber\\
      &-& \frac{1968}{5} g_{BY}^2 g_{BL}^4-64 y_{t}^{\ast} y_{t} g_{BL}^4+672 \lambda_{3} g_{BL}^4-11 \lambda_{3}^3
                - \frac{1024}{5}  \sqrt{\frac{3}{5}} g_{BY}^3 g_{BL}^3-32 \sqrt{15} y_{t}^{\ast} g_{BY} y_{t}g_{BL}^3
   \nonumber\\   
      &+&128  \sqrt{\frac{5}{3}} g_{BY} \lambda_{3} g_{BL}^3+3 g_{2}^2 \lambda_{3}^2
            +\frac{3}{5} g_{BY}^2 \lambda_{3}^2 -36 \lambda \lambda_{3}^2-48 \lambda_{2} \lambda_{3}^2
            -\frac{2139}{25} g_{BY}^4 g_{BL}^2-27 g_{2}^2 g_{BY}^2 g_{BL}^2
   \nonumber\\
       &-& 216 g_{BY}^2 g_{YB}^2 g_{BL}^2+16 \lambda_{3}^2 g_{BL}^2-\frac{228}{5} y_{t}^{\ast} g_{BY}^2 y_{t} g_{BL}^2
         -192 y_{t}^{\ast}g_{YB}^2 y_{t} g_{BL}^2+72 g_{BY}^2 \frac{\lambda }{2} g_{BL}^2
   \nonumber\\
       &+& 48 g_{BY}^2 \lambda_{2} g_{BL}^2+\frac{497}{5} g_{BY}^2 \lambda_{3} g_{BL}^2+1200 g_{YB}^2 \lambda_{3} g_{BL}^2
            +256 \lambda_{2} \lambda_{3} g_{BL}^2-\frac{3}{2} \lambda_{3} Y_{N}^4
   \nonumber\\     
       &-&144 y_{t}^{\ast} g_{YB}^4 y_{t} - \frac{145}{16} g_{2}^4 \lambda_{3}+\frac{1671}{400} g_{BY}^4 \lambda_{3}
              +1512 g_{YB}^3 \lambda_{3} + \frac{9}{8} g_{2}^2 g_{BY}^2 \lambda_{3} 
              -\frac{27}{2} \Big(y_{t}^{\ast}\Big)^2 y_{t}^2 \lambda_{3}
    \nonumber \\
      &-& 15 \lambda ^2 \lambda_{3}-40 \lambda_{2}^2 \lambda_{3}+\frac{17}{4} y_{t}^{\ast} \Big(g_{1}^2
        +g_{BY}^2\Big) y_{t} \lambda_{3}+36 g_{2}^2\lambda  \lambda_{3}+\frac{36}{5} g_{BY}^2 \lambda \lambda_{3}
    \nonumber\\
       &+& 384 g_{YB}^2 \lambda_{2} \lambda_{3}-48 \sqrt{15} y_{t}^{\ast} g_{BY} g_{YB}^2 y_{t} g_{BL}
              +8 \sqrt{\frac{3}{5}} g_{BY}^3 \lambda_{3} g_{BL}+24 \lambda_{3}^2 \Big(g_{YB}^2 + \frac{2}{3} g_{BL}^2\Big)
    \nonumber\\         
       &+& 5 y_{t}^{\ast} y_{t} \lambda_{3} \Big(\frac{9 g_{2}^2}{4}+8 g_{3}^2+g_{YB}^2
            + \frac{2}{3} g_{BL}^2+\sqrt{\frac{5}{2}} g_{1} g_{YB}-\frac{36}{5} \lambda 
             - \frac{12 \lambda_{3}}{5}+\sqrt{\frac{5}{3}} g_{BY} g_{BL}\Big)
     \nonumber \\   
       &+& \frac{1}{4} Y_{N} \Big(\frac{72}{5} g_{1}^2 g_{YB}^2+30 \lambda_{3} g_{YB}^2
            + \frac{48}{5} \sqrt{6} g_{1} g_{BY} g_{BL}  g_{YB}-8 \lambda_{3}^2+\frac{48}{5}g_{BY}^2 g_{BL}^2 
     \nonumber\\     
       &+&  20 \lambda_{3} g_{BL}^2-32 \lambda_{2} \lambda_{3}\Big) Y_{N}.
\eea


\end{document}